\documentclass[a4paper,fleqn,usenatbib]{mnras}

\usepackage{newtxtext,newtxmath}
\usepackage{times}
\usepackage[T1]{fontenc}
\usepackage{ae,aecompl}

\usepackage{graphicx, subfig}
\usepackage[dvipsnames]{xcolor}
\usepackage{xspace}
\newcommand{\ie}{i.\nolinebreak[4]\hspace{0.125em}\nolinebreak[4]e.\@\xspace}
\newcommand{\eg}{e.\nolinebreak[4]\hspace{0.125em}\nolinebreak[4]g.\@\xspace}

\usepackage{amsmath}

\renewcommand{\vec}[1]{\mathbf{#1}}

\newcommand{\Reals}{{\mathbb R}}

\newcommand{\tdim}{d}

\newcommand{\realsource}{``real''\xspace}
\newcommand{\bogussource}{``bogus''\xspace}
\newcommand{\after}{target\xspace}
\newcommand{\before}{template\xspace}
\newcommand{\diff}{difference\xspace}
\newcommand{\code}[1]{{\sc #1}}
\newcommand{\mathup}[1]{\text{\textup{#1}}}
\title[Convolutional Neural Networks for Transient Vetting]{Convolutional Neural Networks for Transient Candidate Vetting in Large-Scale Surveys}

\author[Gieseke et al.]{Fabian Gieseke,$^{1,2}$\thanks{E-mail: fabian.gieseke@di.ku.dk}
Steven Bloemen,$^{3,4}$
Cas van den Bogaard,$^{1}$
Tom Heskes,$^{1}$
\newauthor
Jonas Kindler,$^{5}$
Richard A.~Scalzo,$^{6,7,8}$
Val\'erio A.R.M.~Ribeiro,$^{3,9,10,11}$
Jan van Roestel,$^{3}$
\newauthor
Paul J.~Groot,$^{3}$
Fang Yuan,$^{6,7}$
Anais M\"{o}ller,$^{6,7}$
Brad E.~Tucker,$^{6,7}$
\\
$^{1}$Institute for Computing and Information Sciences, Radboud University, P.O. Box 9010, 6500 GL Nijmegen, The Netherlands\\
$^{2}$Department of Computer Science, University of Copenhagen, Sigurdsgade 41, 2200 Copenhagen, Denmark\\
$^{3}$Department of Astrophysics/IMAPP, Radboud University, P.O. Box 9010, 6500 GL Nijmegen, The Netherlands \\
$^{4}$NOVA Optical InfraRed Instrumentation Group, Oude Hoogeveensedijk 4, 7991 PD Dwingeloo, The Netherlands\\
$^{5}$Institute of Cognitive Science, University of Osnabr\"uck, Wachsbleiche 27, 49090 Osnabr\"uck, Germany\\
$^{6}$Research School of Astronomy and Astrophysics, Australian National University, Canberra, ACT 2611, Australia\\
$^{7}$ARC Centre of Excellence for All-Sky Astrophysics (CAASTRO), Australia\\
$^{8}$Centre for Translational Data Science, University of Sydney, Darlington, NSW 2008, Australia\\
$^{9}$CIDMA, Departamento de F\'isica, Universidade de Aveiro, Campus de Santiago, 3810-193 Aveiro, Portugal \\
$^{10}$Instituto de Telecomunica\c{c}\~oes, Campus de Santiago, 3810-193 Aveiro, Portugal \\
$^{11}$Department of Physics and Astronomy, Botswana International University of Science and Technology, Private Bay 16, Palapye, Botswana\\
}

\date{Accepted XXX. Received YYY; in original form ZZZ}

\pubyear{2017}

\begin{document}
\label{firstpage}
\pagerange{\pageref{firstpage}--\pageref{lastpage}}
\maketitle

\begin{abstract}
Current synoptic sky surveys monitor large areas of the sky to find variable and transient astronomical sources. As the number of detections per night at a single telescope easily exceeds several thousand, current detection pipelines make intensive use of machine learning algorithms to classify the detected objects and to filter out the most interesting candidates. A number of upcoming surveys will produce up to three orders of magnitude more data, which renders high-precision classification systems essential to reduce the manual and, hence, expensive vetting by human experts. We present an approach based on convolutional neural networks to discriminate between true astrophysical sources and artefacts in reference-subtracted optical images. We show that relatively simple networks are already competitive with state-of-the-art systems and that their quality can further be improved via slightly deeper networks and additional preprocessing steps -- eventually yielding models outperforming state-of-the-art systems. In particular, our best model correctly classifies about 97.3\% of all \realsource and 99.7\% of all \bogussource instances on a test set containing 1,942 \bogussource and 227 \realsource instances in total. Furthermore, the networks considered in this work can also successfully classify these objects at hand without relying on difference images, which might pave the way for future detection pipelines not containing image subtraction steps at all.
\end{abstract}


%%%%%%%%%%%%%%%%%%%%%%%%%%%%%%%%%%%%%%%%%%%%%%%%%%

%%%%%%%%%%%%%%%%% BODY OF PAPER %%%%%%%%%%%%%%%%%%

\section{Introduction}

A number of large optical survey telescopes such as
\emph{Skymapper}~\citep{KellerSchmidt2007}, the \emph{Palomar Transient
Factory}~\citep[PTF,][]{RauKulkarni2009}, and
\emph{Pan-STARSS1}~\citep{KaiserBurgett2010} are searching for transient events.
New generation surveys will be able to scan large amounts of the sky faster and deeper allowing
searches for extremely rare or hitherto undiscovered events, such as possible
electromagnetic counterparts of gravitational wave
sources~\citep[see, \eg,][]{NissankeKasliwal2013,SmarttChambers2016}. Those surveys will increase our statistical samples of more
common events, such as supernovae, for experiments in cosmology
and fundamental physics \citep{riess98,schmidt98,scp99}.

The detection of rare transient events among the vast majority of relatively constant sources is an important yet challenging task. Most surveys use difference imaging to find variable stars and transients.
This is usually achieved by performing a pixel-by-pixel subtraction of a pre-existing template image
from the image of interest. Astrophysical sources that are variable or were
absent in the template image remain, while constant sources --
which represent the vast majority of the detected sources -- are removed at
the pixel level. During the difference imaging process, the \before image is
aligned and resampled to take into account distortions in the \after image,
and a convolution is done to match the point-spread function
in all regions of the image \citep[see, \eg,][]{AlardLupton1998,Alard2000}. 

While this process, in principle, allows one to very efficiently find rare
transient events, in practice many of the resulting images contain a large
number of \bogussource~objects. These \bogussource~objects trigger source finding algorithms,
but rather than being of astrophysical nature, they are, in reality,
artefacts. Such artefacts can result from a variety of processes such as issues with image processing (\eg, bad alignment at the subtraction step, between the
\before and \after images), detector imperfections, atmospheric dispersion and cosmic
rays passing through the detector.

The number of potential detections can be very large, with thousands of events
per night produced by current synoptic surveys, and millions of detections per
night expected from future surveys such as the \emph{Large Synoptic Sky
Survey}~\citep[LSST,][]{IvezicTyson2008}. The classification of the detections
as either \realsource or \bogussource sources is a necessary but daunting
task, which will become an even more serious problem in the future. Manual
verification by humans is expensive and, most likely, impossible
to conduct for the amounts of data expected. For this reason, automatic detection
algorithms that yield both a high purity and a high completeness will play an
essential role for future transient surveys.

Machine learning aims at constructing \emph{models} that can perform classification tasks in an automatic manner~\citep{HastieTF2009,Murphy2012}. One particular subfield of machine learning techniques, called \emph{deep learning}~\citep{LeCunBH2015}, has gained considerable attention during the past few years. Deep learning algorithms have successfully been applied to a variety of real-world tasks. Two recent trends have sparked the interest in such algorithms: (1) the dramatic increase of data volumes in almost any field, which, in turn, has produced a massive amount of labelled data that can be used to train and evaluate the models; and (2) the enormous increase in compute power, particularly due to massively-parallel devices such as graphics processing units~(GPUs), which led to a significant reduction of the practical runtime needed to generate deep architectures~\citep{CoatesHWWCN13}.

This paper aims at improving the automatic identification of transient
sources in astronomical images. A standard approach, usually
implemented in current detection pipelines, is based on extracting features
from photometric images, such as the fluxes of the detected sources. Given such a representation of the objects at hand, one resorts to well-established machine learning algorithms. The most widely used approaches are currently based on some kind of dimension reduction (e.g., by conducting a principal component analysis in the preprocessing phase or by extracting physically-motivated features such as magnitudes or parameters that reflect the shape of the point spread function) and a subsequent application of classification methods such as \emph{random forests}, \emph{support vector machines}, \emph{nearest neighbour techniques}, or (standard) \emph{artificial neural networks}~\citep[\eg,][]{BloomRichards2012,2013MNRAS.435.1047B,GoldsteinAndrea2015,MoriiIkeda2016,WrightSmartt2015,duBuissonSivanandam2015}. 
Some recent works also resort to deep neural networks (here, the term ``deep'' refers to the number of hidden layers in a network, see Section~\ref{sec:background} for details). For example, a recent approach resorts to a neural network with three hidden layers that is applied given physically-motivated features~\citep{MoriiIkeda2016}. Another approach is based on \emph{recurrent neural networks} with up to two hidden layers, given time series data that stems from flux values extracted from different observations~\citep{CharnockM2016}. Note that these schemes also resort to an explicit feature extraction step that is conducted in the preprocessing phase.

This paper focuses on improving current detection pipelines and classification systems by means of \emph{convolutional neural networks}~\citep{LeCunBDHHHJ1989,LeCunBH2015}. In contrast to ``standard'' deep architectures, convolutional neural networks do \emph{not} rely on a feature extraction step conducted in the preprocessing phase. Instead, these models ``learn'' good features based on the raw input image data. While convolutional neural networks have already been considered in the context of astronomy \citep{Dieleman21062015,KimB2016}, we present the first application of such models for the task of transient vetting. In this paper we make use of a dataset compiled in the framework of the \emph{Skymapper} supernova searches~\citep{ScalzoETAL2016}. %Our results clearly indicate the potential of convolutional neural networks for the task at hand and we expect such models to be one of the key drivers for the efficient and semi-automatic detection of variable and transient astronomical sources in upcoming surveys. [NOTE: This is really at the conclusion stage]

\section{Background}
\label{sec:background}

In this section, we provide some machine learning background related to the techniques used in this work.
% the state-of-the-art approaches in random forest and convolutional neural networks. The latter is a widely used algorithm in current detection pipelines and in this paper for comparison.

\subsection{Random Forests Revisited}
\begin{figure}
\centering
 \subfloat[Tree 1]{%
  \resizebox{0.28\columnwidth}{!}{\includegraphics{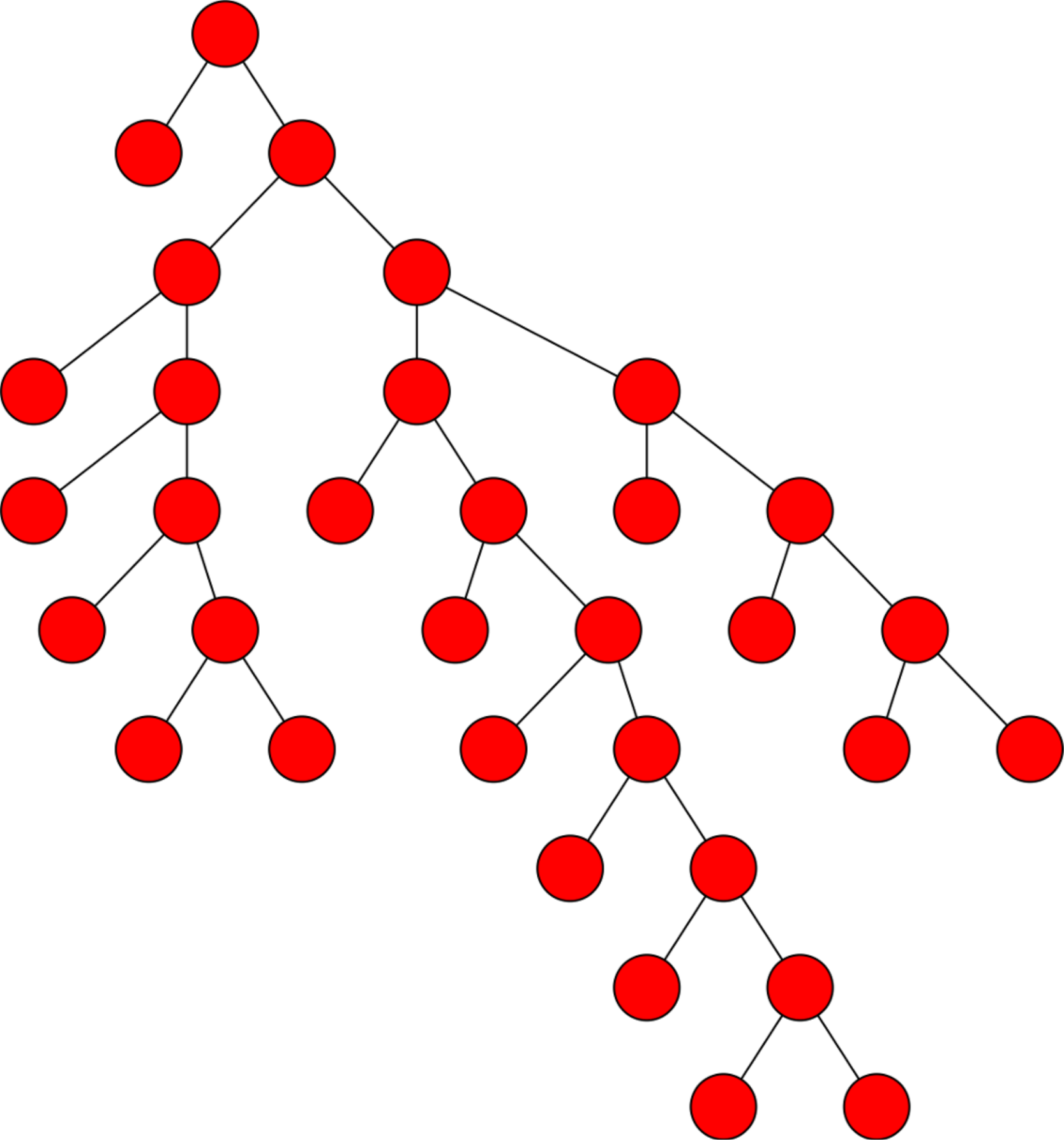}}
 }   	
 \hfill    
 \subfloat[Tree 2]{%
  \resizebox{0.28\columnwidth}{!}{\includegraphics{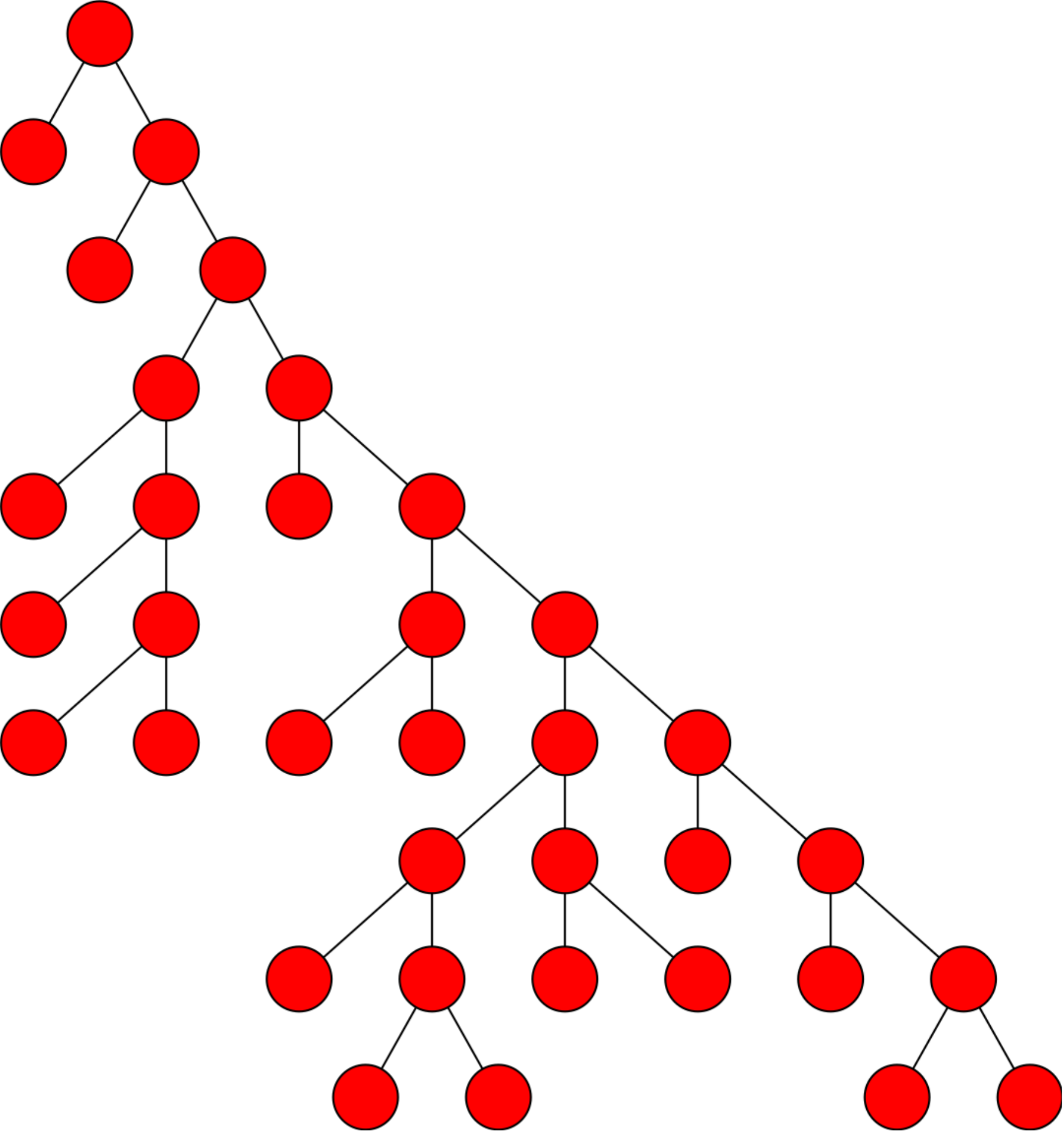}}
 }   	
 \hfill
 \subfloat[Tree 3]{%
  \resizebox{0.28\columnwidth}{!}{\includegraphics{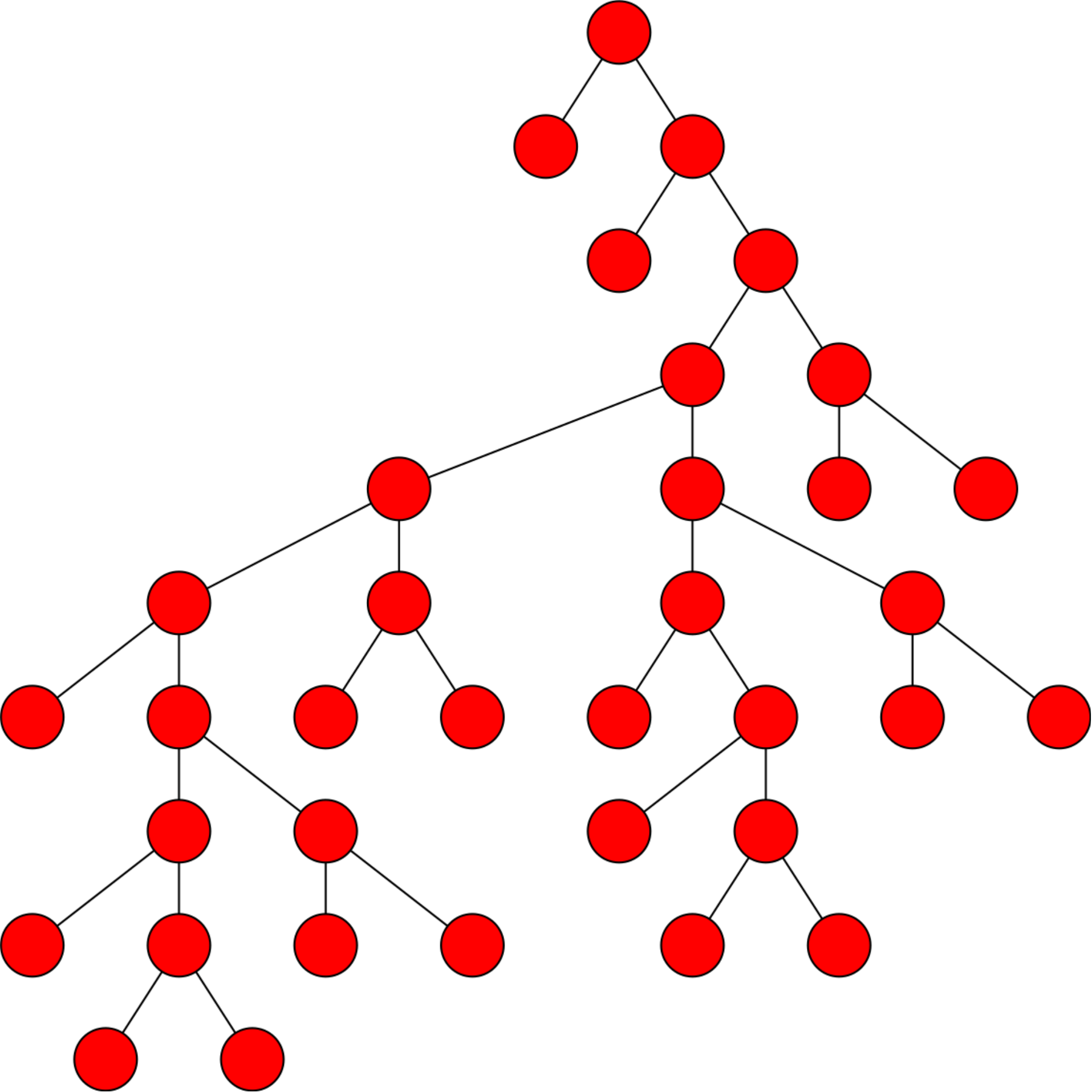}}
 }   	 
 \vskip0.1cm
 \caption{A random forest built for 50 training points. Each tree of
the ensemble is built from top to bottom and at each node, slightly different
``splits'' are used -- resulting in different tree structures. The construction takes place until the leaves are \emph{pure}, meaning that only patterns belonging to the same class are given in a single leaf (resulting in \emph{less} than 50 leaves in this case). For splitting up the nodes, one resorts to different criteria such as the mean squared error for regression scenarios or the Gini index for classification tasks. A random forecasts combines the predictions made by the individual trees.}
\label{fig:forest_example}
\end{figure}

%One of the state-of-the-art approaches used in current detection pipelines (and also in this paper for comparison) is based on \emph{random forests}~\citep{Breiman2001}. 
Random forests depict ensembles of individual classification trees~\citep{Breiman2001,HastieTF2009,Murphy2012}. In general, ensemble methods are among the most successful models in machine learning. This is especially true for random forests, which often yield high accuracies while being, at the same time, conceptually very simple and resilient against small changes of the involved parameter assignments. Since their introduction more than a decade ago, random forests have been extended and modified in various manners. A standard random forest consists of many individual trees (\eg, classification or regression trees), where each tree is built in a slightly ``different'' way (see below) and the ensemble combines the benefits of all of them, see Figure~\ref{fig:forest_example}. 

The trees of a random forest are usually constructed independently from each other. Each tree is built from top to bottom, where the root corresponds to all training instances and the leaves to subsets of the training data. During the construction, each internal node is recursively split into two children such that the resulting subsets exhibit a higher ``purity''. The overall process stops as soon as the leaves are ``pure'' (\ie, they only contain patterns with the same label) or as soon as some other stopping criterion is fulfilled. The original way of building a random forest is based on subsets of the training patterns, one subset for each tree to be built, called \emph{bootstrap samples}~\citep{Breiman2001}. These subsets are drawn uniformly at random (with replacement) to obtain slightly different training sets and, hence, trees. 

The quality of a node split is measured in terms of the gain in \emph{purity}, which, in turn, is measured via different metrics depending on the desired outcome. Typical metrics include, for example, the \emph{mean squared error} for the regression case or the \emph{Gini index} for classification problems~\citep{Breiman2001}. For example, a pure split would be one yielding children containing only instances belonging to the same class, therefore, no further splits are required~\citep{Breiman2001,HastieTF2009,Murphy2012}.

Given a new, unseen instance, one can obtain a prediction for each single tree by traversing the tree from top to bottom based on the splitting information stored in the internal nodes until a leaf node is reached. The labels stored in this leaf are then combined to obtain the prediction for a single tree (\eg, by considering the mean for regression scenarios). The overall prediction of the random forest is based on a combination of the individual predictions. For regression scenarios, one usually simply averages the predictions. For classification settings, one can resort to a majority vote. In summary, the individual trees of a random forest can be seen as ``different'' experts, whose opinions are combined to obtain a single overall prediction for a new instance. 

\subsection{Deep Convolutional Neural Networks}

% We propose a classification approach that is based on \emph{convolutional neural networks}~(CNNs)~\citep{FukushimaM1982,LeCunBH2015,LeCunBDHHHJ1989,RumelhartHW1986}, which are very popular tools in the context of image recognition or natural language processing. As mentioned above, such networks belong to a larger group of so-called deep learning techniques that have gained considerable attention in many application domains over the past few years. While having already been introduced in the late eighties, their current popularity mostly stems from (a) the fact that modern compute architectures can handle very large structures efficiently and at low cost and from (b) the fact that the data volumes have increased dramatically over the past decade in both research and industry. 

Below we briefly introduce convolutional neural networks. For a more detailed description, we refer the reader to the excellent overview by~\citet{LeCunBH2015} and the articles with applications to astronomical research~\citep[e.g.,][]{Dieleman21062015,KimB2016}.

\subsubsection{Artificial Neural Networks}

\begin{figure}
\begin{center}
\resizebox{0.95\columnwidth}{!}{
\resizebox{0.98\textwidth}{!}{\includegraphics{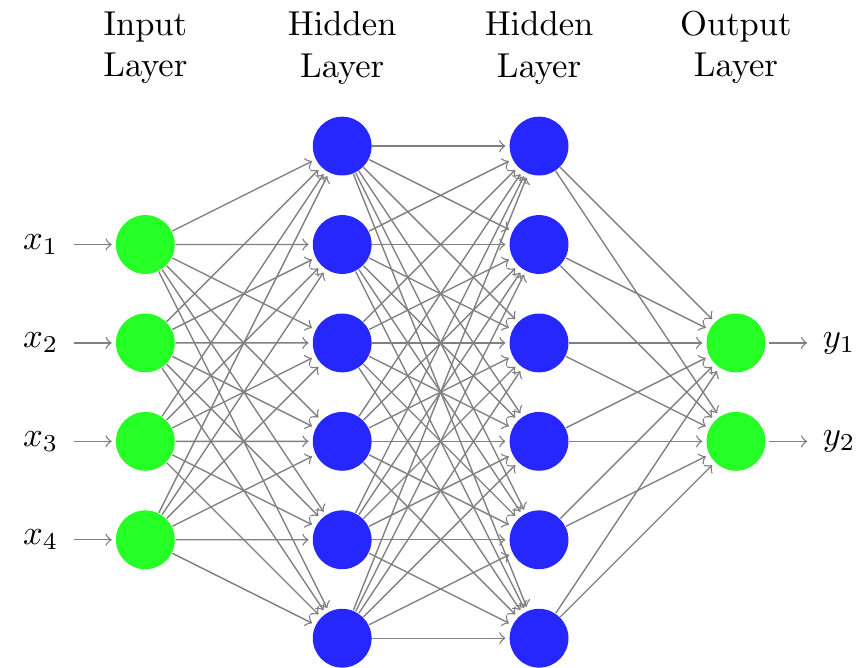}}
}
\end{center}
\vskip0.1cm
 \caption{A standard fully-connected artificial neural network with two hidden layers. The nodes of the network are called neurons and the output of each neuron corresponds to the weighted sum of its input neurons, transformed by an activation function. The output layer contains one neuron for each class and, given an input instance, outputs the corresponding class probabilities (in the case of classification scenarios).}
 \label{fig:anns}
\end{figure}

Convolutional neural networks are special types of the standard \emph{artificial neural networks}~(ANNs)~\citep{HastieTF2009,Murphy2012}, which, in turn, consist of collections of interconnected nodes. 
% These nodes, also called \emph{neurons}, aim at mimicking the behaviour of a human brain and were already invented in the late sixties. 
In a nutshell, a multilayered artificial neural network is based on several \emph{layers}, where the output of a given layer serves as input for the next layer. The first layer is called the \emph{input layer} and the last layer the \emph{output layer}. In between, there requires at least on \emph{hidden layer}. For standard artificial neural networks, these layers are \emph{fully connected}, meaning that all nodes of a given layer are connected to all nodes of the next layer (Figure~\ref{fig:anns}).\footnote{For convolutional neural networks, this is usually not the case except for the last layers (see below).} 

The input layer is specified via the available input data. For example, given a feature vector $\vec{x} \in \Reals^\tdim$ extracted from an image (\eg, a set of magnitudes), each of the feature values $x_1, \ldots, x_\tdim$ corresponds to one of the input nodes of the input layer. Similarly, the output layer is determined by the learning task at hand. For a binary classification problem (\eg, \bogussource vs. \realsource), the output layer consists of two nodes that, for a given input instance, eventually output the class probabilities for each of the classes (note that in this special case, a single output node is sufficient since the classes are mutually exclusive).

The output of the first hidden layer is obtained via the weights $\mathup{W}_1$ associated with the connections between the input layer and the first hidden layer. More specifically, the output of layer $j$ is given by the transformation rule 
% {\bf [sorry, this is a bit confusing, is it the output of the hidden layer rather than the actual output layer? Partly because you use $\vec{x}$ as the input layer but for the output layer you use $\vec{y}$ but my neive perception is that here you are talking about the hidden layers - maybe you should rethink how you have defined $\vec{x}_j$ and specially since you have already said above about the different values in $x_1$... to be the feature values! But I think I see that you mean the you are stepping through the different layers; input --> hidden 1 --> hidden 2 --> output] A potential solution might be to change the definition of $\vec{x}$ since above you called it a feature vector rather than the layer vextor :)}
\begin{equation}
\label{eq:ann}
 \vec{x}_j = f(\mathup{W}_j \vec{x}_{j-1}+ \vec{b}_{j}),
\end{equation}
where $\mathup{W}_{j}$ is the weight matrix associated with the connections between layer $j-1$ and $j$, $\vec{b}_{j}$ a vector containing so-called \emph{bias values} associated with layer $j$, and $f:\Reals^K \rightarrow \Reals^K$ a so-called \emph{activation function} with $K$ being the total number of neurons in layer $j$. Note that the transition from layer $j-1$ to $j$ basically corresponds to applying a standard linear model followed by an element-wise application of the activation function $f$~\citep{HastieTF2009,Murphy2012}. The dimensions of all involved vectors and matrices depend on the number of nodes and connections between the nodes. For example, for the transition from the input to the first hidden layer in Figure~\ref{fig:anns}, we have $\vec{x}_0 \in \Reals^4$, $\vec{x}_1 \in \Reals^6$, $\mathup{W}_1 \in \Reals^{6 \times 4}$, and $\vec{b}_1 \in \Reals^6$. Popular choices for the activation function $f:\Reals^K \rightarrow \Reals^K$ are
\begin{itemize}
 \item the \emph{linear activation function} ${[f(\vec{x})]}_i=x_i$,
 \item the \emph{rectified activation function} ${[f(\vec{x})]}_i = \max(0, x_i)$,
 \item or the \emph{softmax activation function} ${[f(\vec{x})]}_i=\frac{e^{x_i}}{\sum_{j=1}^{K} e^{x_j}}$
\end{itemize}
for $i=1,\ldots,K$. Hence, the layers of a neural network iteratively transform the feature representation $\vec{x} \in\Reals^\tdim$ of an input instance. Finally, the vector at the last hidden layer is transformed to only a single output neuron for standard regression scenarios or to multiple output neurons for classification or multivariate regression scenarios. Note that the different layers can resort to different activation functions, except for the last one, the output layer, which is somewhat restricted by the learning task at hand. For regression tasks, one usually makes use of a linear activation function, whereas the softmax activation function is a common choice for classification scenarios. Hence, given a new instance $\vec{x}\in\Reals^\tdim$ for which one would like to obtain a prediction (\eg, if it is of type \bogussource or \realsource), one consecutively applies the transformation rule (Equation~\ref{eq:ann}), which eventually yields the output vector $\vec{y}$. For regression scenarios, $\vec{y}\in\Reals^1$ corresponds to the prediction made by the network for the input $\vec{x}$, whereas the vector $\vec{y}\in\Reals^C$ contains the class probabilities given classification scenarios with $C$ possible classes. 

Training such a neural network basically corresponds to finding weights such that the network performs well on new, unseen data (\eg, fewer misclassifications). Various optimization techniques can be applied (\eg, variants of gradient descent) 
% {\bf [This is the first time you mention this and sounds like you assume the reader will know what gradient descent it since you delve right in to the ``variants'' - prehaps you should mention briefly what gradient descent are and how this is applied to finding the weights]} 
so that the output of the network becomes more consistent with the class labels given for the training data. The so-called \emph{learning rate} $\gamma>0$ is a parameter used by many of the underlying optimisation techniques that affects the size of the weight updates (\eg, similar to the step-size of standard gradient descent). The learning rate is a parameter that needs to be specified beforehand (as the network structure itself) and is usually tuned via grid search (\ie, various assignments are tested and one resorts to the training data to evaluate the induced quality). 

Another important parameter is the number of \emph{epochs}, which usually correspond to a full pass over the available training data or to processing a certain batch of a fixed size of training instances (most optimisation techniques process the training instances iteratively). For the latter option, the \emph{batch size} determines the number of instances processed per single epoch~\citep{HastieTF2009}. 
% For details related to the training and optimization process, we refer to \cite{HastieTF2009} and references therein.

\subsubsection{Convolutional Neural Networks}

In recent years, so-called \emph{deep networks} have become more and more popular. Here, the term ``deep'' is related to the number of hidden layers. A special type of such deep architectures are convolutional neural networks, which consist of multiple layers of different types. Such networks have been successfully applied in the context of many application domains. We will focus on image-based input data for the description of convolutional neural networks.

A typical convolutional neural network with multiple hidden layers is shown in Figure~\ref{fig:cnn}. As standard artificial networks, convolutional neural networks also exhibit an input and an output layer. Furthermore, the last hidden layers often correspond to standard fully-connected dense layers as well. In contrast, the first hidden layers are conceptually very different and consist of various types of layers. The most prominent types, applied in this work, are \emph{convolutional layers}, \emph{pooling layers}, and \emph{dropout layers}:
\begin{figure*}
\begin{center}
\resizebox{0.75\textwidth}{0.35\textheight}{\includegraphics{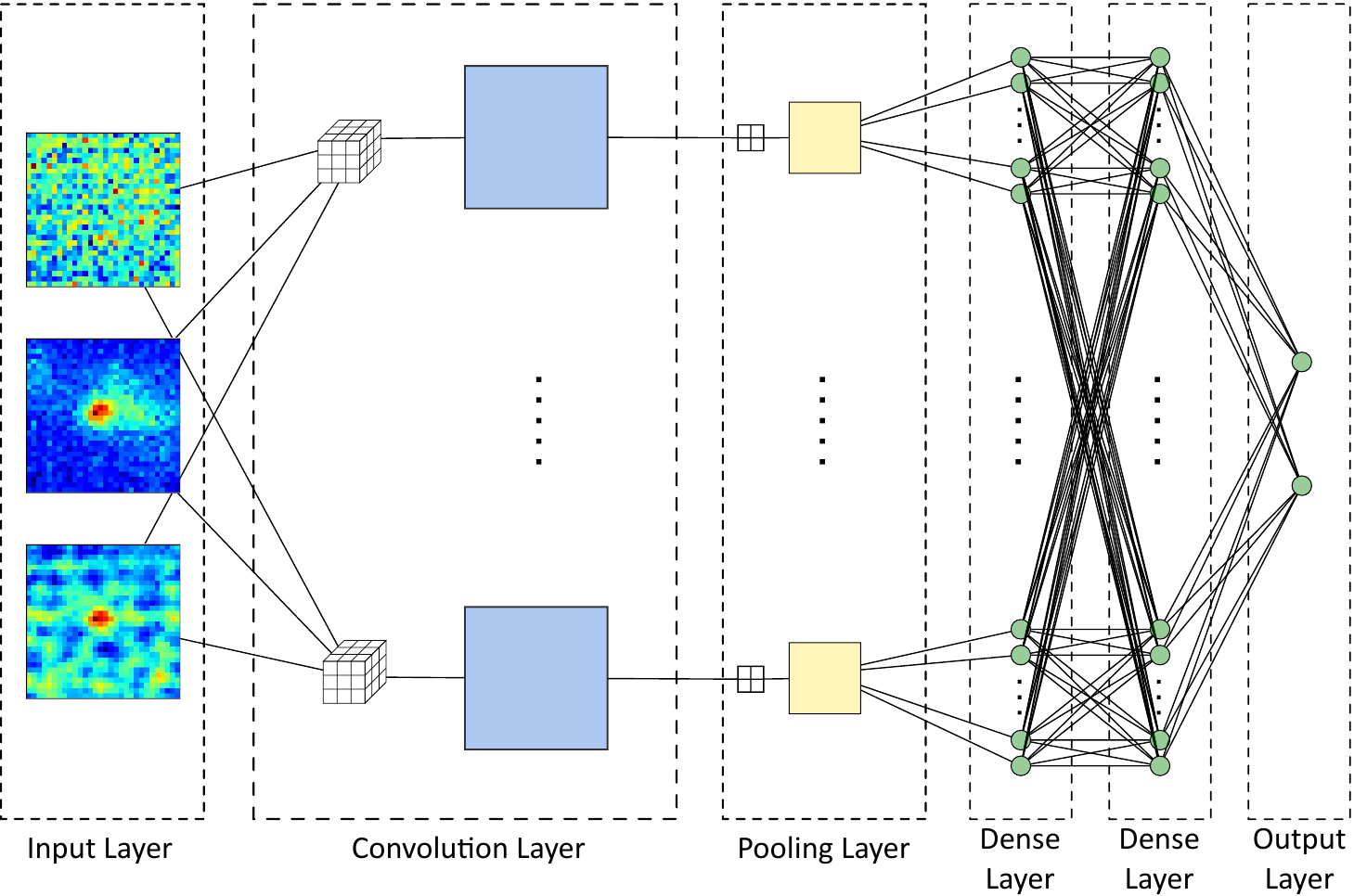}}
\end{center}
\caption{A convolutional neural network with one convolutional and one pooling layer, followed by two fully connected standard hidden layers and an output layer. Prior to the fully-connected hidden layers, the pixel-based feature maps are flattened, meaning that all pixel values of all feature maps of the previous layers are concatenated to form a single vector. Generally, by resorting to multiple convolutional and pooling layers, convolutional neural networks are capable of learning a hierarchical feature representation of the input instances -- starting with simple features at early layers and more complex features towards the end.}
\label{fig:cnn}
\end{figure*}
\begin{itemize}
 \item \emph{Convolutional layers:} Such layers form the basis for convolutional neural networks and yield so-called \emph{feature maps} by sliding a small window of weights across the input feature maps that stem from the previous layer (the input images of the input layer form the initial feature maps). In a nutshell, each feature map of a given layer stems from convolving all input feature maps using a set of \emph{filters} (weight matrices), one filter for each input feature map. For example, in Figure~\ref{fig:cnn}, we have, for each feature map in the first hidden layer, three filters of size $3 \times 3$ that are used to convolve all three input feature maps. The sum of all these convolved images correspond to the term $\mathup{W}_{j} \vec{x}_{j-1}$ in equation~(\ref{eq:ann}). Afterwards, a matrix of bias values is added to this sum image, followed by an element-wise application of an activation function~$f$.
 \item \emph{Pooling layers:} This is the second prominent type of layers. Pooling layers are used to decrease the number of learnable parameters (the filters/weight matrices). More precisely, such a pooling layer reduces the sizes of the feature maps by aggregating pixel values. For example, a max-pooling layer considers patches within each feature map and replaces each patch by the maximum value in that patch (\eg, in Figure~\ref{fig:cnn}, each feature map of the previous layer is processed via patches of size $2 \times 2$, leading to new feature maps of half the size). Naturally, other operations can be applied such as taking the mean of a given pixel patch.
 \item \emph{Dropout layers:} These guard against \emph{overfitting}, in which the trained network relies heavily on aspects of the training data that do not generalise well to new, unseen data. Dropout layers randomly omit hidden units by setting their value to zero with a user-defined probability $p \in (0,1)$ such that other hidden units cannot fully rely on them (such techniques are also used in standard artificial neural networks). Thus, these layers force the network to rely on more units, preventing a reliance on noise or artefacts (dropout layers are not shown in Figure~\ref{fig:cnn} since dropout basically only affects the underlying optimisation process).
\end{itemize}

Therefore, convolutional layers aim to extract features that are somewhat invariant against translation. Note that there are also significantly less weights/connections for such convolutional layers (only the values given in the filters/weight matrices have to be learnt). Similarly, pooling layers make the network invariant against small transitions and reduce the number of nodes in the network (\ie, pixels in the feature maps). These two modifications often significantly increase the classification performance of such networks compared to standard fully-connected artificial neural networks.

\section{Deep Transient Detection}

In this section we provide details of the different models considered for this paper in order to detect transients. 

\subsection{Imaging Data}
\label{subsec:photo_data}
\begin{figure*}
\subfloat[Bogus]{
\begin{minipage}{0.99\columnwidth}
\resizebox{0.98\textwidth}{!}{\includegraphics{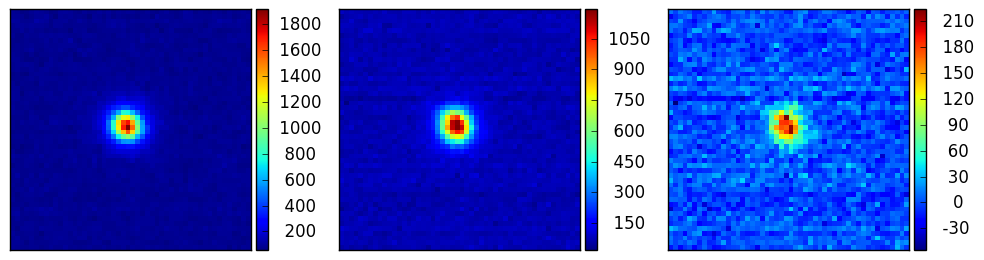}}
\resizebox{0.98\textwidth}{!}{\includegraphics{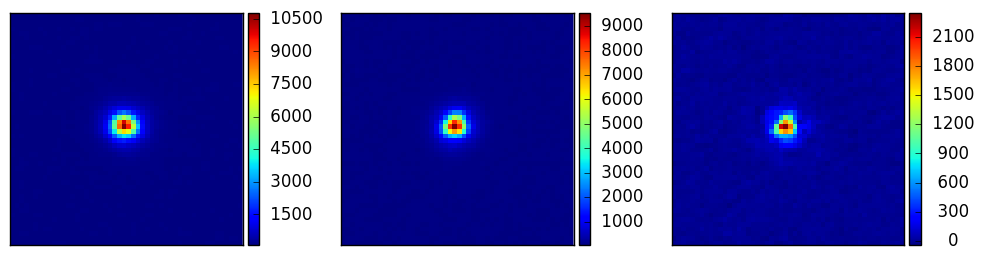}}
\resizebox{0.98\textwidth}{!}{\includegraphics{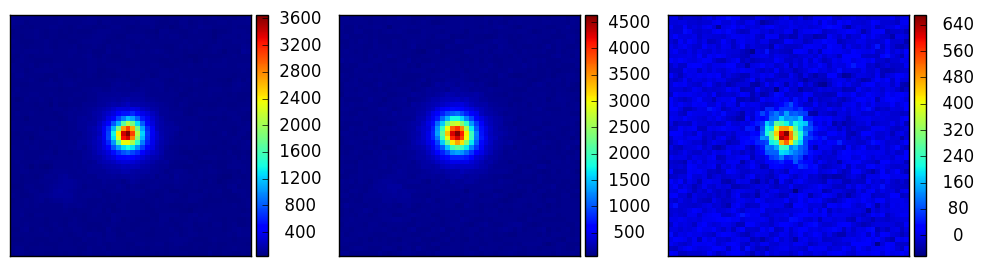}}
\end{minipage}
}
\hfill
\subfloat[Real]{%
\begin{minipage}{0.99\columnwidth}
\resizebox{0.98\textwidth}{!}{\includegraphics{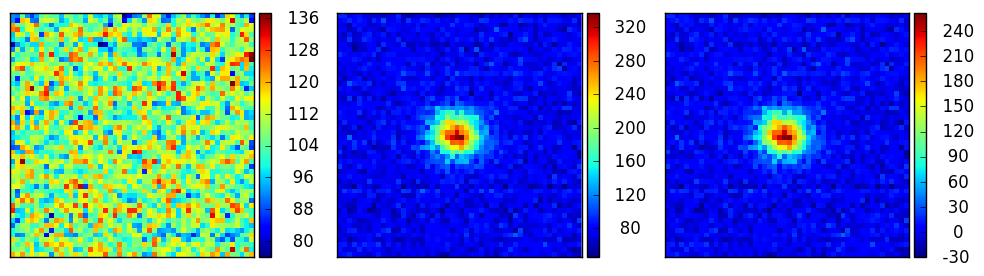}}
\resizebox{0.98\textwidth}{!}{\includegraphics{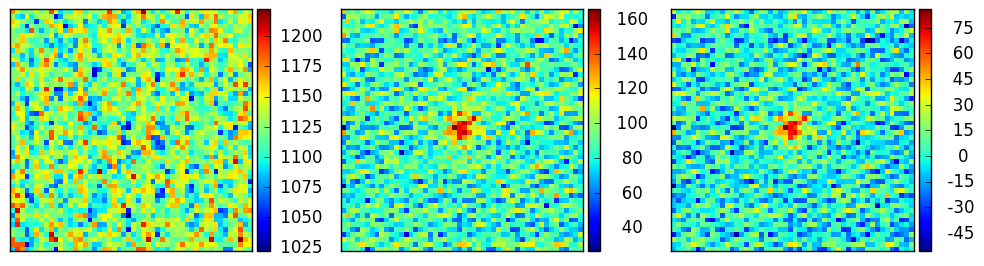}}
\resizebox{0.98\textwidth}{!}{\includegraphics{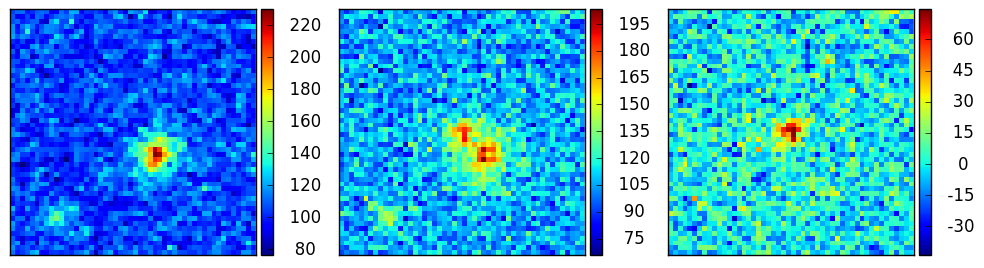}}
\end{minipage}
}
\caption{``Best'' case examples of three \bogussource and three \realsource. The columns represent the \before,
\after, and \diff images per field (from left to right) and the rows represent different fields. The different colours along with the colour bars illustrate the pixel intensities per image. For \realsource sources, there is usually no flux in the centre on the \before image.
% {\bf [I am not sure this add much to the discussions - since I imagine that the \after image will set the bounding box - to which the \before image is then bound, and since the \after image will have the object in the centre then we shouldn't see anything in the \before image. Actually, I am, also, not sure that the rest of the sentences after this are relevant - other the very last sentence which is important.]} 
However, there may be ``misleading'' instances very close to the centre of the image (which, ideally, would no longer be present in the \diff image). The majority of the instances in the dataset are simple cases. As shown in our experiments, there is still a significant number of difficult instances outstanding, which can be very challenging for machine learning models to identify correctly. The unit of the colour scale is in counts.}
\label{fig:examples}
% \vspace{-15pt}
\end{figure*}

\begin{table*}
 \caption{Features used as input for the random forest models.} %{\bf [I removed the word photometric because not everything here is photometric]}
 \label{tab:features}
 \begin{tabular}{ll}
  \hline
  Feature & Description\\
  \hline
  \texttt{xsub}    & x coordinate on difference image, in pixels \\
  \texttt{ysub}    & y coordinate on difference image, in pixels \\
  \texttt{esub}    & ellipticity of source on difference image \\
  \texttt{thsub}   & direction of semi-major axis of source on difference image \\
  \texttt{fwhmsub} & full width at half maximum of all difference image sources \\
  \texttt{f4sub}   & flux within 4-pixel aperture in difference image \\
  \texttt{f8sub}   & flux within 8-pixel aperture in difference image \\
  \texttt{flagsub} & \code{sextractor} source flags in difference image \\
  \texttt{starsub} & \code{sextractor} star-galaxy score in difference image \\
  \texttt{xref}    & x coordinate on template image, in pixels \\
  \texttt{yref}    & x coordinate on template image, in pixels \\
  \texttt{eref}    & ellipticity of source on template image \\
  \texttt{thref}   & semi-major axis of source on template image \\
  \texttt{fwhmref} & full width at half maximum of all template image sources \\
  \texttt{f4ref}   & flux within 4-pixel aperture in template image \\
  \texttt{flagref} & \code{sextractor} source flags in template image \\
  \texttt{starref} & \code{sextractor} star-galaxy score in template image \\
  \texttt{enew}    & ellipticity of source on difference image \\
  \texttt{thnew}   & semi-major axis of source on difference image \\
  \texttt{fwhmnew} & full width at half maximum of all target image sources \\
  \texttt{f4new}   & flux within 4-pixel aperture in target image \\
  \texttt{flagnew} & \code{sextractor} source flags in target image \\
  \texttt{starnew} & \code{sextractor} star-galaxy score in target image \\
  \texttt{n2sig3}  & number of at least 2-sigma negative pixels
                     in 3x3 box in difference image \\
  \texttt{n3sig3}  & number of at least 3-sigma negative pixels
                     in 3x3 box in difference image \\
  \texttt{n2sig5}  & number of at least 2-sigma negative pixels
                     in 5x5 box in difference image \\
  \texttt{n3sig5}  & number of at least 3-sigma negative pixels
                     in 5x5 box in difference image \\
  \texttt{nmask}   & number of masked pixels in 5x5 box in the target image \\
  \texttt{Rfwhm}   & fwhmnew/fwhmref ratio \\
  \texttt{goodcn}  & surface density of detected sources on subtraction \\
  \texttt{subconv} & direction of convolution
                     (template-target or target-template) \\
  \texttt{nndref}  & distance in pixels to nearest neighbour source
                     in template image \\
  \texttt{nndnew}  & distance in pixels to nearest neighbour source
                     in target image \\
  \texttt{apsig4}  & signal-to-noise ratio of 4-pixel aperture flux
                     in difference image \\
  \texttt{apsig8}  & signal-to-noise ratio of 4-pixel aperture flux
                     in difference image   \\
  \texttt{normrms} & ratio of square root of isophotal area
                     in difference image to fwhmsub \\
  \texttt{normfwhm} & ratio of full width at half maximum
                     in difference image to fwhmsub \\
  \texttt{Rfref}   & signal-to-noise ratio of nearest counterpart
                     in difference image \\
  \texttt{Raref}   & ratio of candidate semi-major axis to all sources
                     in difference image \\
  \texttt{Reref}   & ratio of candidate ellipticity to all sources
                     in difference image \\
  \texttt{Dthref}  & difference in candidate semi-major axis direction
                     from all sources in difference image \\
  \texttt{Rfnew}   & signal-to-noise ratio of nearest counterpart
                     in target image \\
  \texttt{Ranew}   & ratio of candidate semi-major axis to all sources
                     in target image \\
  \texttt{Renew}   & ratio of candidate ellipticity to all sources
                     in target image \\
  \texttt{Dthnew}  & target in candidate semi-major axis direction
                     from all sources in target image \\
  \hline
 \end{tabular}
\end{table*}
Our models were trained on early science operations, prior to April 2015, imaging data from the \emph{Skymapper Supernova and Transient Survey}.
The difference imaging pipeline is described in more detail in~\citep{ScalzoETAL2016}. The main image processing tools used are the 
\code{SWarp}~\citep{swarp} for astrometric registration and resampling
to a common coordinate system,
\code{SExtractor}~\citep{sextractor} for source detection and photometry,
and \code{HOTPANTS}\footnote{{http://www.astro.washington.edu/users/becker/v2.0/hotpants.html}}
for photometric registration and image subtraction.
Data for each example include the \emph{\before} image,
the \emph{\after} image,
and the \emph{\diff} image. 
% {\bf [I was not sure if it was relevant when the images were taken, I think people will understand the \before and \after idea]}
In Figure~\ref{fig:examples} we show examples of both \realsource and \bogussource taken from the dataset.

% NET1
%         elif net_type == "astronet7_32_64":
% 
%             layers7 = [
%                 (InputLayer, {'shape': (None, X.shape[1], X.shape[2], X.shape[3])}),
%             
%                 (Conv2DLayer, {'num_filters': 32, 'filter_size': (3, 3), 'pad': 0}),
%                 (MaxPool2DLayer, {'pool_size': (2, 2)}),
%                 (DropoutLayer, {'p':0.1}),
%                                             
%                 (DenseLayer, {'num_units': 64}),
%                 (DropoutLayer, {'p':0.5}),
%                 (DenseLayer, {'num_units': 64}),
%             
%                 (DenseLayer, {'num_units': len(set(y)), 'nonlinearity': softmax}),
%             ]
% 
%             net = NeuralNet(
%                 layers=layers7,
%                 max_epochs=epochs,
%                 update=lasagne.updates.adam,
%                 update_learning_rate=learning_rate,
%                 objective_l2=0.0025,
%                 train_split=TrainSplit(eval_size=0.05),
%                 verbose=verbose,
%             )     
%             
%             return net  
\begin{table*}
\subfloat[Net1(A,B)]{
\begin{minipage}{0.32\textwidth}
\centering
  \begin{tabular}{lll}\hline
      Type & Size & Parameters \\ \hline
      input & $3 \times 30 \times 30$ & \\
      conv & $A \times 28 \times 28$ & fs=(3,3)\\
      maxpool & $A \times 14 \times 14$ & ps=(2,2)\\
      dropout & $A \times 14 \times 14$ & p=0.1\\
      dense & $B$ & \\
      dropout & $B$ & p=0.5\\
      dense & $B$ & \\
      dense & $2$ & \\ \hline
  \end{tabular}
\end{minipage}
}
\hfill
% NET2
%         elif net_type == "astronet6":
% 
%             layers6 = [
%                 (InputLayer, {'shape': (None, X.shape[1], X.shape[2], X.shape[3])}),
%             
%                 (Conv2DLayer, {'num_filters': 32, 'filter_size': (3, 3), 'pad': 0}),
%                 (MaxPool2DLayer, {'pool_size': (2, 2)}),
%                 (DropoutLayer, {'p':0.1}),
% 
%                 (Conv2DLayer, {'num_filters': 128, 'filter_size': (3, 3), 'pad': 0}),
%                 (MaxPool2DLayer, {'pool_size': (2, 2)}),
%                 (DropoutLayer, {'p':0.1}),
%                                                             
%                 (DenseLayer, {'num_units': 512}),
%                 (DropoutLayer, {'p':0.5}),
%                 (DenseLayer, {'num_units': 512}),
%             
%                 (DenseLayer, {'num_units': len(set(y)), 'nonlinearity': softmax}),
%             ]
% 
%             net = NeuralNet(
%                 layers=layers6,
%                 max_epochs=epochs,
%                 update=lasagne.updates.adam,
%                 update_learning_rate=learning_rate,
%                 objective_l2=0.0025,
%                 train_split=TrainSplit(eval_size=0.05),
%                 verbose=verbose,
%             )   
%
%             return net  
\subfloat[Net2]{%
\begin{minipage}{0.32\textwidth}
\centering
  \begin{tabular}{lll}\hline
      Type & Size & Parameters \\ \hline
      input & $3 \times 30 \times30$ & \\
      conv & $32 \times 28 \times 28$ & fs=(3,3)\\
      maxpool & $32 \times 14 \times 14$ & ps=(2,2)\\
      dropout & $32 \times 14 \times 14$ & p=0.1\\
      conv & $128 \times 12 \times 12$ & fs=(3,3)\\
      maxpool & $128 \times 6 \times 6$ & ps=(2,2)\\
      dropout & $128 \times 6 \times 6$ & p=0.1\\      
      dense & $512$ & \\
      dropout & $512$ & p=0.5\\
      dense & $512$ & \\
      dense & $2$ & \\ \hline
  \end{tabular}
\end{minipage}
}
\hfill
% NET 3
%         elif net_type == "astronet2":
% 
%             layers2 = [
%                 (InputLayer, {'shape': (None, X.shape[1], X.shape[2], X.shape[3])}),
%             
%                 (Conv2DLayer, {'num_filters': 16, 'filter_size': (3, 3), 'pad': 0}),
%                 (MaxPool2DLayer, {'pool_size': (2, 2)}),
%                 (DropoutLayer, {'p':0.1}),
% 
%                 (Conv2DLayer, {'num_filters': 32, 'filter_size': (3, 3), 'pad': 0}),
%                 (MaxPool2DLayer, {'pool_size': (2, 2)}),
%                 (DropoutLayer, {'p':0.3}),
%                 
%                 (Conv2DLayer, {'num_filters': 64, 'filter_size': (3, 3), 'pad': 0}),
%                 (MaxPool2DLayer, {'pool_size': (2, 2)}),
%                 (DropoutLayer, {'p':0.5}),                
%             
%                 (DenseLayer, {'num_units': 1000}),
%                 (DropoutLayer, {'p':0.5}),
%                 (DenseLayer, {'num_units': 1000}),
%             
%                 (DenseLayer, {'num_units': len(set(y)), 'nonlinearity': softmax}),
%             ]
% 
%             net = NeuralNet(
%                 layers=layers2,
%                 max_epochs=epochs,
%                 update=lasagne.updates.adam,
%                 update_learning_rate=learning_rate,
%                 objective_l2=0.0025,
%                 train_split=TrainSplit(eval_size=0.05),
%                 verbose=verbose,
%             )            
%             
%             return net
\subfloat[Net3]{%
\begin{minipage}{0.32\textwidth}
\centering
  \begin{tabular}{lll}\hline
      Type & Size & Parameters \\ \hline
      input & $3 \times 30 \times30$ & \\
      conv & $16 \times 28 \times 28$ & fs=(3,3)\\
      maxpool & $16 \times 14 \times 14$ & ps=(2,2)\\
      dropout & $16 \times 14 \times 14$ & p=0.1\\
      conv & $32 \times 12 \times 12$ & fs=(3,3)\\
      maxpool & $32 \times 6 \times 6$ & ps=(2,2)\\
      dropout & $32 \times 6 \times 6$ & p=0.1\\      
      conv & $64 \times 4 \times 4$ & fs=(3,3)\\
      maxpool & $64 \times 2 \times 2$ & ps=(2,2)\\
      dropout & $64 \times 2 \times 2$ & p=0.1\\            
      dense & $1000$ & \\
      dropout & $1000$ & p=0.5\\
      dense & $1000$ & \\
      dense & $2$ & \\ \hline
  \end{tabular}
\end{minipage}
}
\caption{Network structures considered in this work (`fs' denotes the filter size of the convolutional layer, `ps' the pooling size of the pooling layer and `p' the dropout probability). A and B are parameters determining the input sizes of the layers.
}
\label{tab:cnns}
\end{table*}
We trained our random forest models on features extracted from the images (Table~\ref{tab:features}), while the convolutional neural networks
were applied directly on the images themselves. To evaluate the performance of each model, we split the available data into a
training set and a test set of roughly equal size.

To make the most of our limited number of \realsource training
instances, we enabled training on multiple distinct detections of the same
transient candidate on different nights; this allowed us to sample more real
variation in seeing and sky
background level for the same candidate than we would if we included only
one detection of each candidate.  Because we expected the classifier results
to also be affected by details of the host galaxy placement and morphology,
which would be the same for multiple observations of the same transient,
we placed all observations of the same object (\eg, a supernova)
within the same partition (\ie, training or test); this choice enables our
training methodology to make honest estimates of the generalization error
to entirely new transient sources with different host galaxy properties.
Apart from this constraint, the partition each individual transient candidate
occupied was chosen at random.
The training set contains 2,237 instances
(2,010 \bogussource and 227 \realsource)
of 851 distinct \bogussource and 140 distinct \realsource sky objects.
The test set contains 2,236 instances
(2,009 \bogussource and 227 \realsource)
of 851 distinct \bogussource and 141 distinct \realsource sky objects.

% The training set contains 2,237 instances (2,010 \bogussource and 227 \realsource)
% and the test set 2,236 (2,009 \bogussource and 227 \realsource).
% The split between training and test data was made in such a way that multiple observations of the same object were always added to the same partition (\ie, training or test).
% Hence, this partition into training and test instances mimics the realistic scenario in which classification
% systems have to be able to detect newly discovered
% sources. 
% {\bf [Which split, the 50/50 split or the partition of the data. If it is the former I am not sure I see why that is the case. If it is the partition maybe use a different word than ``split''.]} 
We removed those instances that were located less than 15 pixels from the edge of a CCD, in both the training and test sets, since the pixel cut-outs we use for our analysis (see Section \ref{sec:network_structures}) would be incomplete in these cases. This yields a final training set with 2,162 instances (1,939 \bogussource and 223 \realsource) and a test set containing 2,169 instances (1,942 \bogussource and 227 \realsource).

\subsection{Baseline: Random Forests \& Features}
\label{susec:random_forest_features}

For the use of random forests, we extract various features from the imaging data (see Table~\ref{tab:features}).
Most of the features are taken from \citet{BloomRichards2012}.  
% {\bf [You've already said this earlier, feels like dejavu! Perhaps delete and use:]} 
A large fraction of the features reflect
properties of the source in the \diff image, as well as some contextual
information such as the presence of, and distance from, a nearby neighbour in
the \before image \citep{BloomRichards2012}. We supplemented the features above with new features intended to capture the global properties of
the images - these flag obviously bad (\eg, trailed) images
and subtractions. We also included \code{SExtractor} error codes and
star-galaxy separation scores; the latter provide a redundant output from
a different method (neural network) that may capture aspects of point sources
not covered by our existing features. 
% We will use these features to train the random forest algorithm. Such algorithms are already employed in current pipelines (for example, in the SkyMapper transient survey) and exhibit a very good classification performance. {\bf NOTE: The last sentence feels like you have already said it earlier in the paper. [It might be instructful if in Table~\ref{tab:features} we highlight the actual features used maybe with a little star next to the feature name.]}

As with many other machine learning models, the classification performance depends on the particular parameter assignments used to generate the random forest. However, random forests are usually very robust against small modifications, \ie, given reasonable parameter assignments, the validation performance is often very similar. The main parameters that need to be tuned are: (1) the number of estimators, (2) the number of features tested per split, (3) if bootstrap samples are used or not, and (4) the stopping criterion used. Furthermore, variations of classical random forests exist such as the \emph{extremely randomised trees}~\citep{GeurtsEW2006}, which consider ``random'' thresholds as potential splitting candidates. 

For our analysis, we have tested different random forest variants and
parameter assignments. However, for simplicity, we only report results of a 
single configuration (all others yielded very similar performances). In 
particular, we consider the Gini index to measure the impurity of the internal 
node splits, make use of 500 trees built using a bootstrap sample, resort to 
fully-grown trees, and test $\sqrt \tdim$ features per node split, where 
$\tdim$ is the number of overall features considered.\footnote{We use 
\texttt{Python 2.7} and the \texttt{scikit-learn} package 
(version $0.18$)~\citep{scikit-learn} for processing and analysing the data. 
More precisely, we make use of the 
\texttt{sklearn.ensemble.RandomForestClassifier} class as the random forest 
implementation and initialise the model using the following parameters: 
\texttt{bootstrap=True}, \texttt{n\_estimators=500}, 
\texttt{min\_samples\_split=2}, \texttt{criterion="gini"}, and 
\texttt{max\_features="sqrt"}.} We also tested various other parameter 
assignments, which all yielded very similar classification accuracies (as long 
as a sufficiently large amount of trees was considered).

% We refer to Appendix~\ref{appendix:implementation} for details related to the parameters and the software packages used.

\subsection{Network Structures \& Parameters}
\label{sec:network_structures}
While convolutional neural networks have been successfully applied to several real-world tasks~\citep[see, \eg,][]{LeCunBH2015}, choosing the best-performing network structure (w.r.t. the classification performance on the test set) is often based on trial-and-error. Very deep structures might be disadvantageous given ``simple tasks''. On the other hand, too simplistic structures might not be able to adapt to the learning task at hand and, thus, may yield unsatisfactory results as well. Therefore, the goal is to consider models that are complex enough to capture the characteristics of a learning task and that are, at the same time, not too complex. This is related to the so-called \emph{bias-variance tradeoff}~\citep{HastieTF2009}, which describes the well-known problem in machine learning of finding models with the right complexity such that they perform well on unseen data, and to the optimisation process itself. For example, deeper models generally exhibit more model parameters that need to be tuned/fitted, whereas in shallower networks one usually tunes less parameters. In this paper, we tackled this problem by starting with very simple and shallow convolutional networks and then increasing the complexity of the networks step-wise adding further convolutional layers. As will be shown in our experimental evaluation, simple convolutional neural networks already yield very promising classification results. Furthermore, due to their simplicity, one gains insight into how and why these networks perform so well. Additionally, the performance may be improved by means of data augmentation and further preprocessing steps.

The network structures considered in this paper are presented in Table~\ref{tab:cnns}. Unless stated otherwise, the input layers correspond to the three input images that are available (\ie, \before, \after, \diff). Each network contains at least one max-pooling, convolutional, and dropout layers. The final layers are fully connected, followed by a softmax activation function to obtain class probabilities (\bogussource vs. \realsource). The corresponding parameters for each layer are provided in the table.
% (here, `fs' specifies the \emph{filter size} and `ps' the \emph{pool size} of a convolutional and pooling layer, respectively {\bf [you don't need this bracket since you don't use these definitions again in the text, maybe just leave it in the table]})
We considered 1,000 training iterations for all networks without any data augmentation (see below) and 5,000 for the ones with data augmentation. Here, a training iteration refers to processing a batch of 128 training images.

% Unless stated otherwise, we consider $30\times30$ pixel image cut-outs for all experiments. To train and evaluate the networks, we resort to the \texttt{nolearn} implementation~\citep{nolearn}. 
For the convolutional neural network approaches, we cropped the images to a size of $30 \times 30$ pixels and made use of the \texttt{Python} package \texttt{nolearn} (version 0.6.1.dev0).\footnote{The \texttt{nolearn} package implements various wrappers for the \texttt{Lasagne} package, which, in turn, depicts a lightweight library for the well-known \texttt{Theano} package \citep[see,][for details]{nolearn,lasagne,2016arXiv160502688short}.} More precisely, we made use of the \texttt{nolearn.lasagne.NeuralNet} class and initialised the models with different parameters and layers (see Table~\ref{tab:cnns}). We complemented Net2 and Net3 with data augmentation steps that were conducted on-the-fly per training iteration (i.e., per batch of 128 training instances). In particular, we rotated each image by $90^\circ$, $95^\circ$, $100^\circ$, $180^\circ$, $185^\circ$, $190^\circ$, $270^\circ$, $275^\circ$, and $280^\circ$. Subsequently, all \realsource instances were flipped horizontally and vertically. Note that we did not apply any translation augmentation step since it is guaranteed that all \realsource instances are centred (up to 1 or 2 pixels).\footnote{We also conducted some experiments with very small translation steps, which, however, did not lead to an improvement w.r.t. the classification performance.} Resampling of the augumented images was done in a manner to perserve the flux using the \texttt{scipy.ndimage.interpolation.rotate} function. Since the data augumentation step essentially yields a significantly larger dataset, we increased the number of training iterations (5,000 instead of 1,000). For all networks, we resorted to the \texttt{nolearn} default settings to initialise the weights of all layers (i.e., Glorot with uniformly sampled weights)~\citep{AISTATS2010_GlorotB10}. Furthermore, to train the networks, we resorted to Adam updates with learning rate $\gamma=0.0002$~\citep{journals/corr/KingmaB14}. The overall process aimed to minimise the categorical cross entropy as objective with L2 regularisation (\texttt{objective\_lambda2=0.025}).

% \begin{itemize}
%  \item We fix the learning rate (see Section~\ref{sec:background}) to $\gamma=0.0002$ for all networks (this seemed to be a good choice for all networks and experiments; similar assignments yielded similar results).
% \end{itemize}

We made use of standard low-cost gaming graphics processing units (GPUs), such as \texttt{Nvidia GeForce GTX 770}, to speed up the training and validating processes. Training the models was the most time-consuming phase of the two processes. Here, each training iteration (i.e., processing a batch of 128 images) took about 0.5 to 5 seconds depending on the network architecture and the particular GPU being used. Note, however, that the network models considered in this work can all be generated in a couple of minutes (e.g., 1000 training iterations for the shallow networks) or hours (e.g., 5000 training iterations and the deeper networks). 
% {\bf [We have already made note about what GPUs we use] - why is there a big jump from a couple of minutes to hours if there is only a factor of 5 increase in the sample size? Is it an N$^2$ problem - and then it is only a factor of 25 increase?}

% \begin{itemize}
%  \item \todo Details related to nolearn, lasagne, Theano, stopping criterion (epochs!), GPU was used, a training iteration of size 128 images (called epoch in nolean) took about 0.5 seconds (for the shallow networks) and up to 5 seconds for the deep networks using a standard gaming GPU (here: Nvidia GeForce GTX 770).
% \end{itemize}

% , see Appendix~\ref{appendix:implementation} for details. 

\section{Analysis}

We compared the performance of the random forests approach, currently used in data processing pipelines, with those of different convolutional neural networks described above.

\subsection{Experimental Setup}

All experiments resorted to data described in Section~\ref{subsec:photo_data} to generate and evaluate the models. In the following, we assume that the label for a \realsource instance is $+1$ and the one for a \bogussource instance is $-1$. We split the data into a ``training part'', used for generating the corresponding models, and a ``testing part'', used for the final evaluation of the models' classification performances. We also shuffle the training dataset prior to training the different models. Note that none of the final test instances are shown to the model during the training phase. Therefore, the results indicate the performance of the classifiers on new, unseen data (given the same distribution of objects). 

For fitting the random forest model, we resort to all instances given in the training set. For the convolutional networks, we use 95 percent of the (potentially augmented) training set to actually fit the models, whereas five percent are used as a holdout validation set to monitor the objective values. It is worth mentioning that, for all networks considered, the objective values steadily descreased during the fitting process on both these parts of the training set up to a certain point, where additional epochs did not lead to any significant changes anymore. Furthermore, the objectives on both subsets were very close to each other, indicating that the convolutional neural networks did not overfit on the training set. Hence, both the employed regularization as well as the dropout layers seem to be effective measures against overfitting in this context.

Our datasets are inbalanced with regards to more \bogussource than \realsource instances (approximately one \realsource instance out of 10 instances). The inbalance is relevant when assessing the performance of a given model. A classifier that simply assigns the class \bogussource to all instances might already achieve very high accuracy, is, however, usually not helpful from a practical point of view since all \realsource instances would be missed. For this reason, it is crucial to consider an evaluation criteria that takes such class inbalances into account. The evaluation criteria considered are based on different types of correct and incorrect classifications:

\begin{itemize}
 \item \emph{True positives (tp)}: Is defined as the number of \realsource instances that are correctly classified as \realsource.
 \item \emph{False positives (fp)}: Is defined as the number of \bogussource instances that are misclassified as \realsource.
 \item \emph{True negatives (tn)}: Is defined as the number of \bogussource instances that are correctly classified as \bogussource.
 \item \emph{False negatives (fn)}: Is defined as the number of \realsource instances that are misclassified as \bogussource.
\end{itemize}

The commonly used \emph{accuracy} of a classifier is then given by $(tp+tn)\times(tp+tn+fp+fn)^{-1}$. Further, the \emph{purity} (also called \emph{precision}) is given by $tp \times (tp+fp)^{-1}$ and the \emph{completeness} (also called \emph{recall}) by $tp \times (tp+fn)^{-1}$. Another measure that combines the above numbers into a single number (still being meaningful for unbalanced data) is the so-called \emph{Matthews correlation coefficient}~(MCC)~\citep{Matthews1975}:
\begin{equation}
 \frac{tp\times tn - fp \times fn}{\sqrt{((tp+fp)(tp+fn)(tn+fp)(tn+fn))}}
\end{equation}
Here, an MCC of $+$1.0 corresponds to perfect predictions, $-$1.0 to total disagreement between predictions and the true classes, and 0.0 to a ``random guess''. 

\begin{figure*}
  \centering
    \subfloat[Random Forest]{%
      \resizebox{0.14\textwidth}{!}{\includegraphics{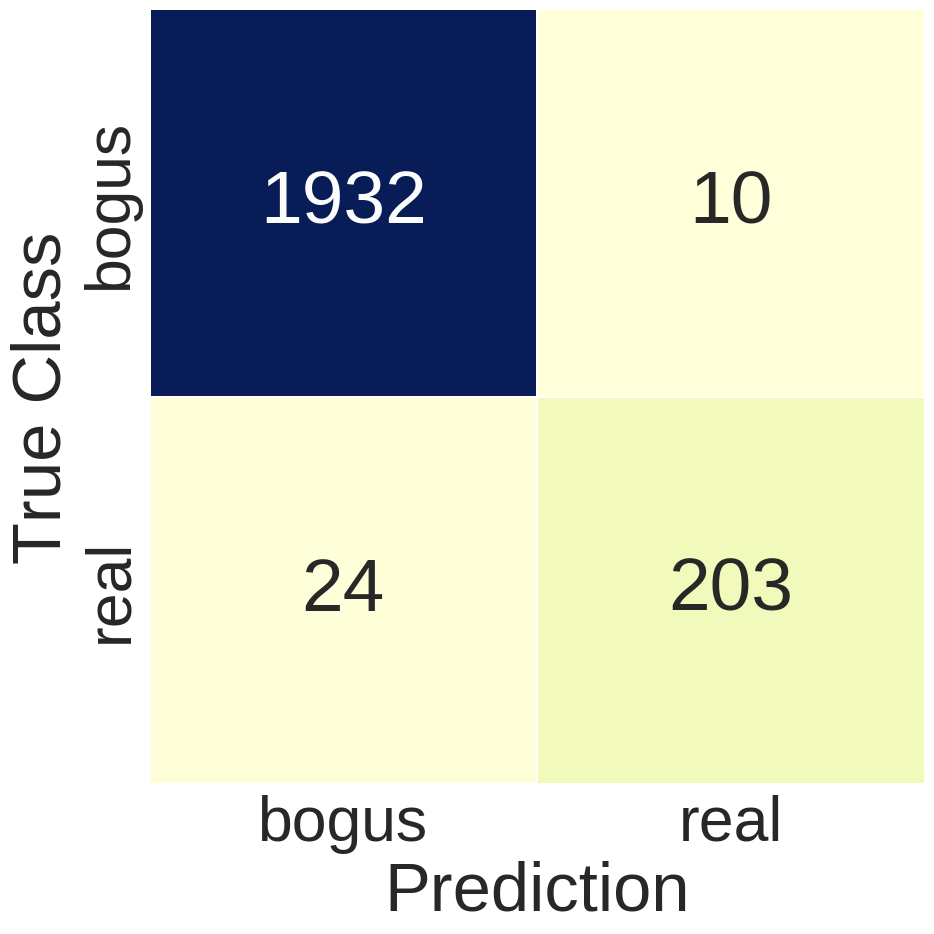}}
    }    
    \hfill
    \subfloat[Net1(32,64)]{%
      \resizebox{0.14\textwidth}{!}{\includegraphics{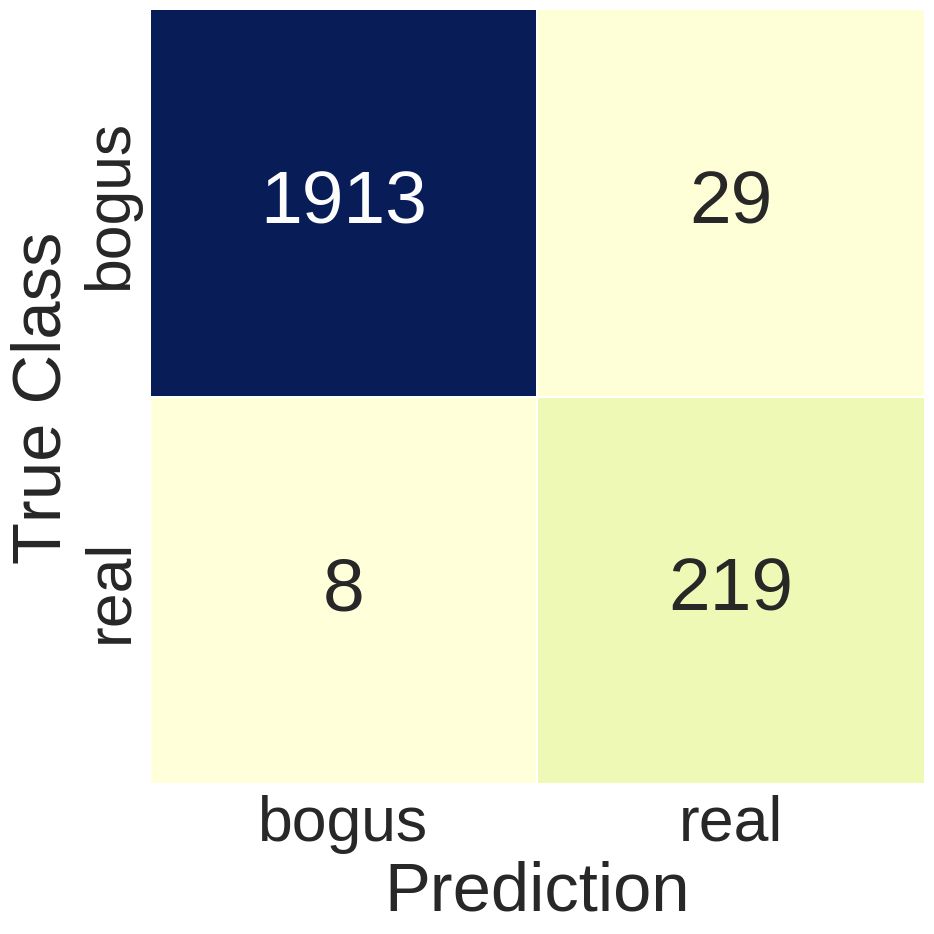}}
    }
    \hfill    
    \subfloat[Net1(64,128)]{%
      \resizebox{0.14\textwidth}{!}{\includegraphics{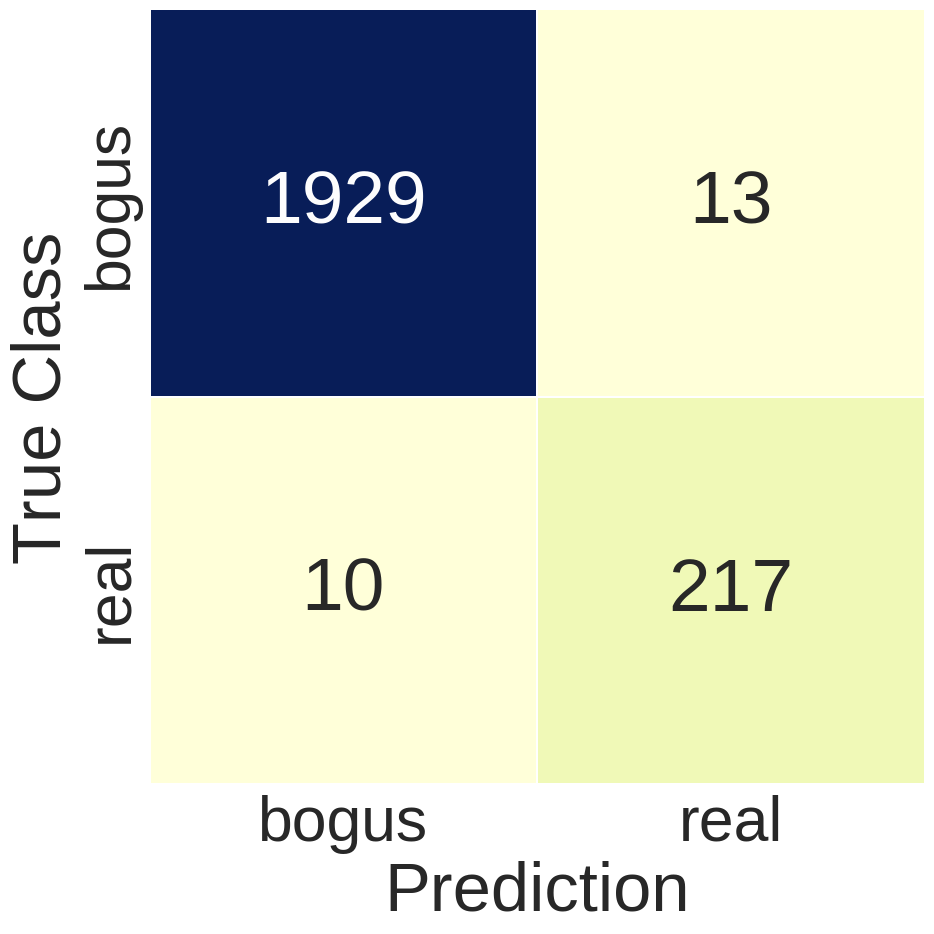}}
    }
    \hfill    
    \subfloat[Net1(128,256)]{%
      \resizebox{0.14\textwidth}{!}{\includegraphics{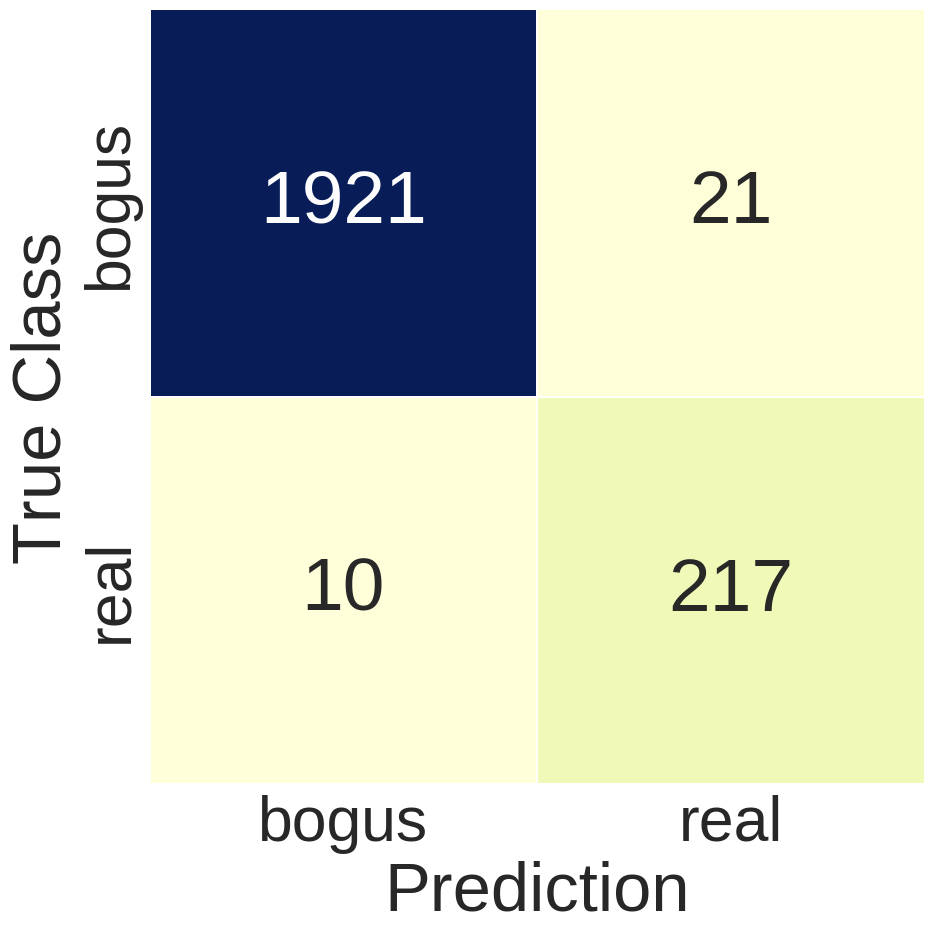}}
    }    
    \hfill
    \subfloat[Net2]{%
      \resizebox{0.14\textwidth}{!}{\includegraphics{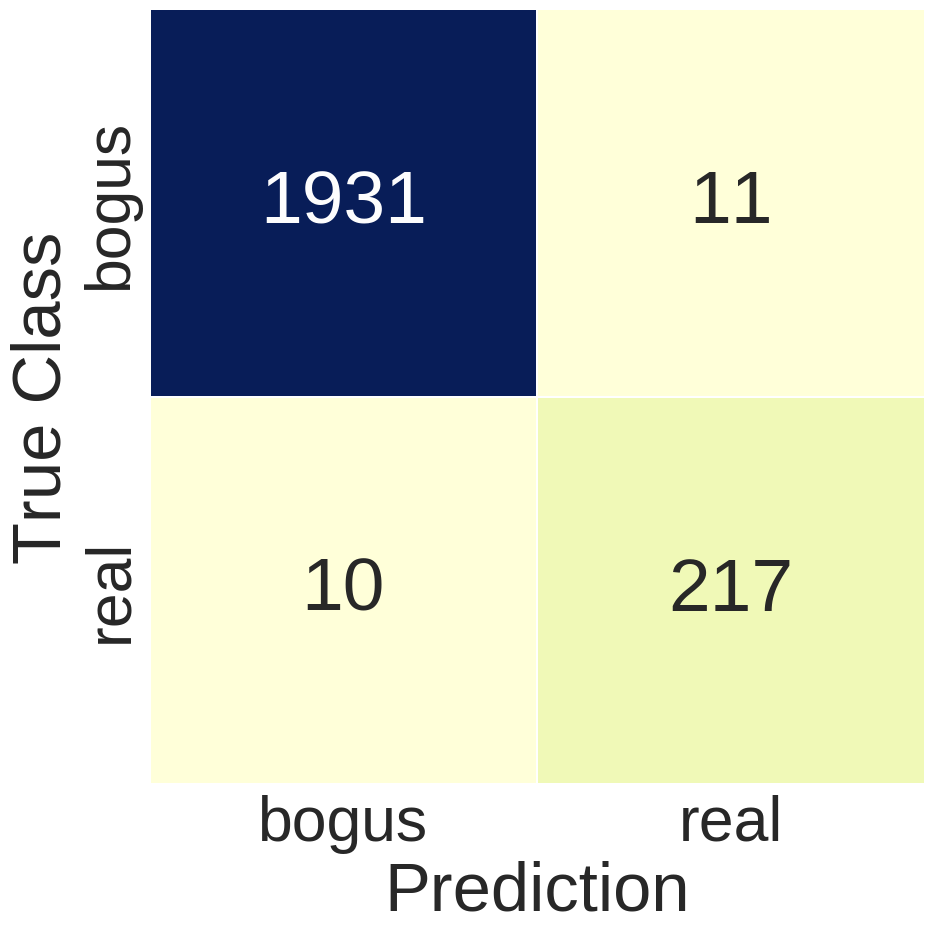}}
    }    
    \hfill    
    \subfloat[Net3]{%
      \resizebox{0.14\textwidth}{!}{\includegraphics{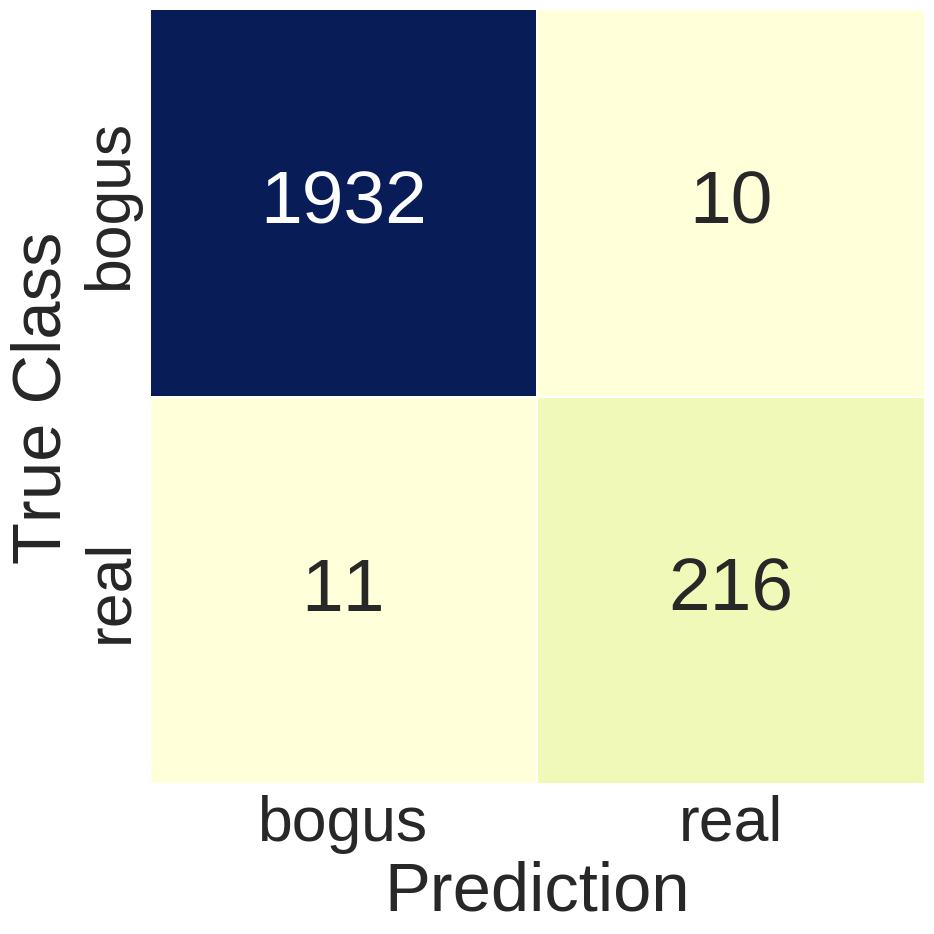}}
    }        
    \caption{Confusion matrices for each of the considered models. While the random forest already achieves a good classification performance, the different networks, also the simple ones, seem to yield competitive or even slightly better results. 
%     {\bf [Shouldn't the values in the confusion matrix be in percent form? That way we can see better the differences.]}
    }
    \label{fig:confusion}  
\end{figure*}
For a new instance, all models considered in this work output some kind of class probability value. More precisely, we consider softmax non-linearities for all convolutional neural networks and the mean predicted class probabilities of all trees for the random forest model (here, the class probability for a single tree is simply the fraction of the samples that belong to that class in a leaf). Unless stated otherwise, we will resort to the default ``threshold'' of $0.5$ to decide for a class label; we will analyse the influence of other thresholds at the end of our experimental evaluation.

From a practical point of view, training convolutional neural networks often takes much longer than training other machine learning models such as random forests. However, these models usually only need to be generated from time to time if new training data becomes available. Computing predictions for new, unseen instances takes considerably less time. This renders the networks described in this work applicable in the context of upcoming surveys with corresponding pipelines processing hundreds of thousands of candidate sources per night.

\subsection{Results}

We start by providing an overall comparison of various classification models and will subsequently analyse some of the models along with certain modifications in more detail.

\subsubsection{Classification Performance}

A meaningful evaluation measure (also for unbalanced data) are confusion matrices containing both the number of correctly classified instances as well as the number of misclassifications. The confusion matrices for various models, obtained via the test set, are shown in Figure~\ref{fig:confusion}. It can be seen that each method performs reasonably well, but some of the models yield a significantly smaller number of misclassifications. In particular, one can make the following observations:

\begin{itemize}
 \item \emph{Random Forest:} The standard random forest performs reasonably well, which indicates that the features extracted from the \diff images capture the main characteristics of the learning task. In total, 34 instances are misclassified out of all 2,169 instances, which corresponds to an MCC of $0.915$. Note that a similar performance can be obtained by slightly different random forest models that stem from different splitting mechanisms or parameter assignments (see Section~\ref{susec:random_forest_features}). However, none of these variants yielded an MCC better than $0.925$.
 \item \emph{Shallow Networks:} The conceptually very simple and shallow networks with only a single convolutional layer already yield a surprisingly good performance that is competitive with the one of the random forest. The MCCs for the Net1(32,64), Net1(64,128), and Net1(128,256) are $0.937$, $0.951$ and $0.933$, respectively. We will investigate these networks and their classification performances in more detail below.
 \item \emph{Deeper Networks:} The deeper networks with two or three convolutional layers yield a slightly better classification performance than the shallow networks. Applying data augmentation steps prior to training the networks seem to further reduce the number of misclassifications (see below). The MCC for Net2 and Net3 without any further data augmentation and transformation steps are $0.949$ and $0.954$, respectively. Net3 only misclassifies 21 out of the 2,169 instances and, hence, 13 less than the random forest (in particular, it misses less \realsource~sources that are assigned to the \bogussource~class). 
%  {\bf [actually one of the cool things is that it, supposidely, the objects that were missclassified in the random forest were real, so it may appear that CNN improves at least what the real rate is. If that makes sense!] :D}
\end{itemize}

It is worth mentioning that the classification performance is, in general, very similar for slightly different convolutional neural networks, \ie, neither the particular network structure nor the involved parameters seem to have a significant influence on the final classification performance. The conclusion one can draw at this point is that standard ``out-of-the-box'' convolutional neural networks seem to be well-suited for the task at hand and that even relatively simple networks yield a performance that is competitive with state-of-the-art approaches.

\subsubsection{Shallow Networks}
\begin{figure}
	\centering
    \subfloat[\bogussource]{%
      \resizebox{0.22\textwidth}{!}{\includegraphics{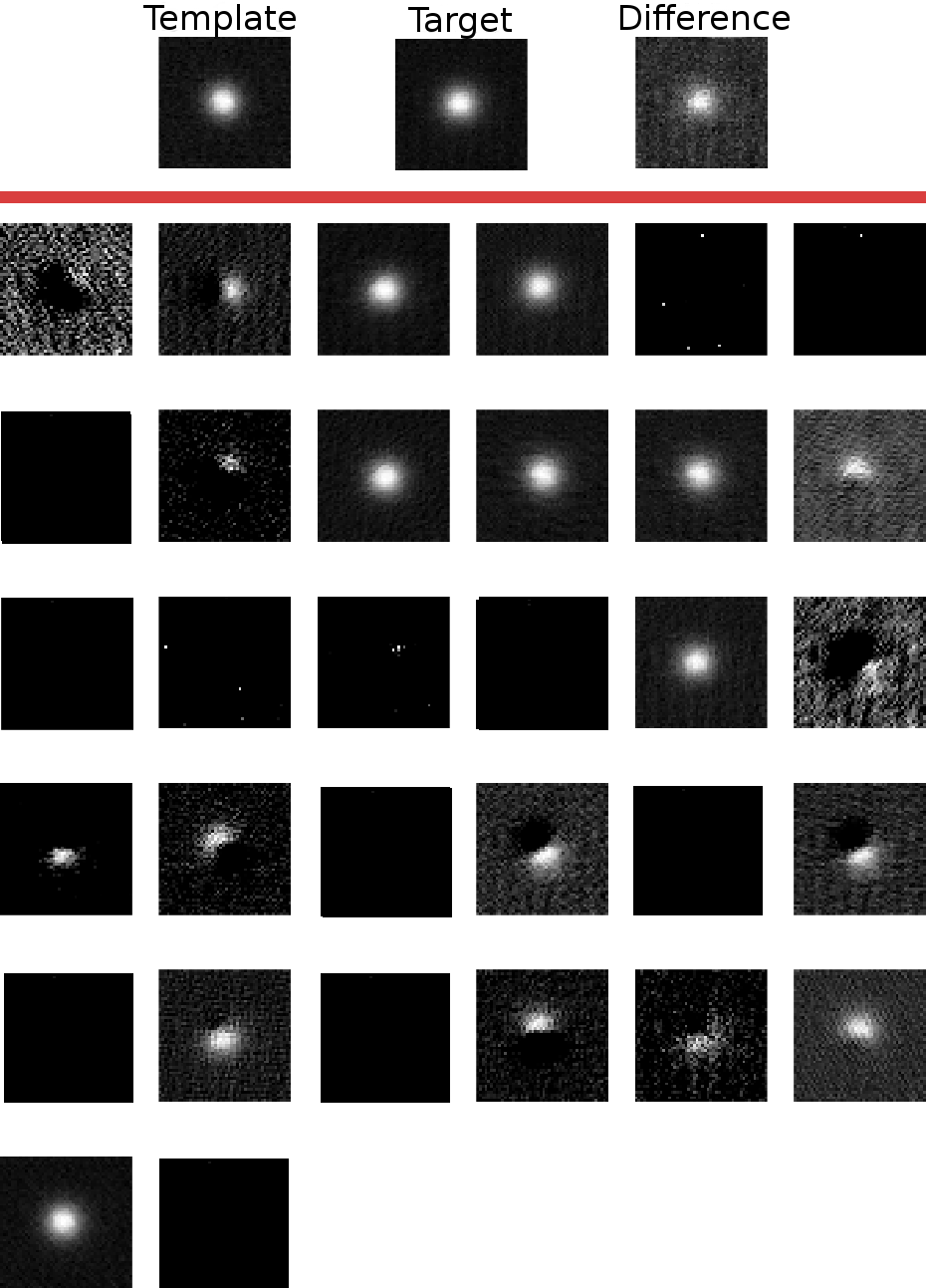}}
    }   	
    \hfill    
    \subfloat[\realsource]{%
      \resizebox{0.22\textwidth}{!}{\includegraphics{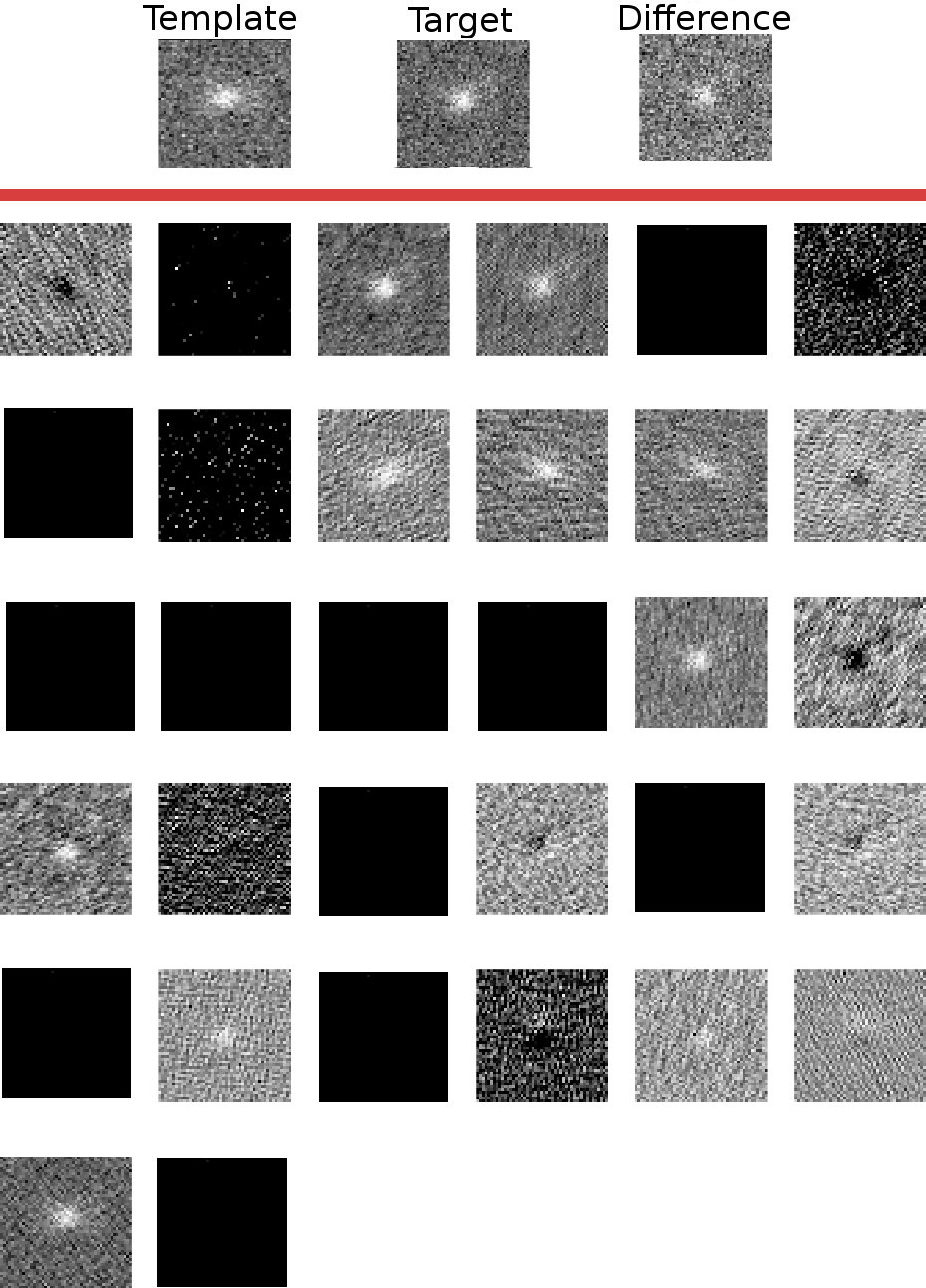}}
    }   	 	
  \caption{Activations (feature maps) of the convolutional layer of Net1(32,64) given a \bogussource and \realsource source, respectively. The top row shows the three input images that are available for each instance. Below the red line, the 32 feature maps are provided that are induced by the convolutional layer of Net1(32,64).}
  \label{fig:activations}
\end{figure}

% \begin{figure*}
% 	\centering
%  	\resizebox{0.3\textwidth}{!}{\includegraphics{images/experiments/convolutions/before.png}}
%  	\hfill
%  	\resizebox{0.3\textwidth}{!}{\includegraphics{images/experiments/convolutions/after.png}}
%  	\hfill
%  	\resizebox{0.3\textwidth}{!}{\includegraphics{images/experiments/convolutions/diff.png}}
% 	\caption{Weights for the \before, \after, and \diff images (left to right) for the first convolutional layer of Net1(32,64). For each weight matrix, the sum of all weights is provided on top of the associated image.}
% 	\label{fig:weights}
% \end{figure*}
Although being conceptually very simple, the shallow networks already yield a very good classification performance. This is actually surprising since the images are obtained under \emph{different} observational conditions (such as the phases of the moon) that will lead to background levels that differ from image to image even though the same region/object is observed. Intuitively, ``subtracting'' this background level in the preprocessing phase should simplify the learning problem and is, for this reason, also part of other approaches~\citep{2013MNRAS.435.1047B}. However, we observed that such a normalisation step in the preprocessing phase actually \emph{decreases} the classification performances of the networks, both for the shallow and deeper models.

\begin{figure}
	\centering
 	\resizebox{0.44\textwidth}{!}{\includegraphics{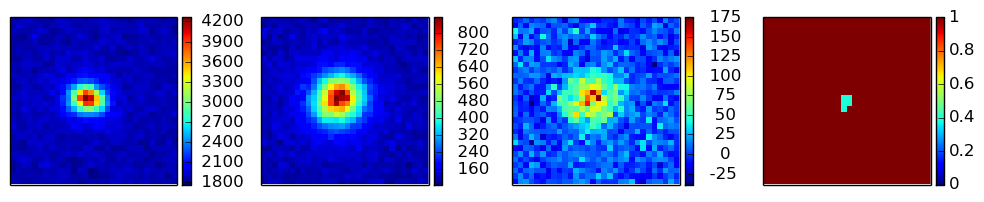}}
 	\resizebox{0.44\textwidth}{!}{\includegraphics{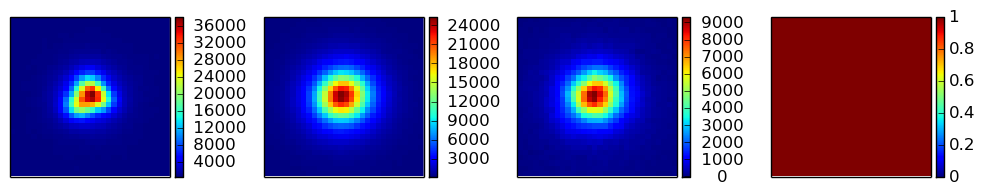}}
 	\resizebox{0.44\textwidth}{!}{\includegraphics{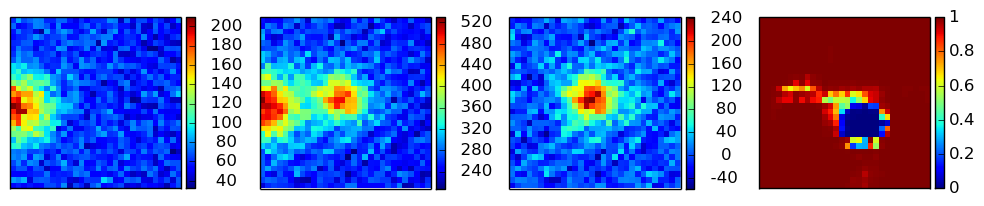}}
 	\resizebox{0.44\textwidth}{!}{\includegraphics{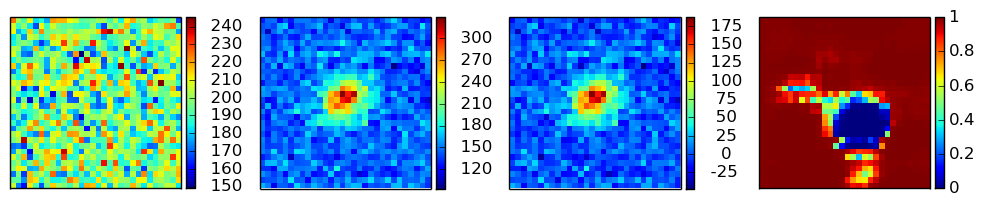}}
	\caption{Input images along with occlusion maps (right column) for two \bogussource (top) and two \realsource (bottom) sources. The different colours along with the colour bars illustrate the pixel intensities for each of the three left images per row, whereas they sketch the different probabilities for the occlusion map in the rightmost image per row. For the \realsource sources, the centres of the input images seem to be important, whereas for the \bogussource instances, obscuring any part of the image appears to have little effect.}
	\label{fig:occlusion}
\end{figure}

To investigate why the shallow networks already perform so well, we consider the simplest architecture, Net1(32,64), which only contains a single convolutional layer. In Figure~\ref{fig:activations}, the activations of the feature maps induced by a \bogussource and a \realsource instance are shown. In both instances, there appears to be feature maps that activate on the background (``dark centre''), while other maps activate on different parts of the centre. This suggests that the network can distinguish between the source itself and the background, thus being able to classify images relatively unhindered by different levels of noise. 
% Further insights might also be gained by analysing the weights of the single convolutional layer of Net1(32,64), which are provided in Figure~\ref{fig:weights}.
% for the single convolutional layer of Net1(32,64). 
% The weights indicate that different feature maps focus on different combinations of the input images (\eg, some of the weight matrices for the \before image only contain values close to zero). 
% Another 
% An interesting observation is the fact that, for a large number of feature maps, the sum of all weights given in a single weight matrix is close to zero as well (\eg, \before image, top row, fourth matrix). This indicates that the network has actually learned to automatically ``subtract'' the background level since applying such a convolution filter to a region containing just noise is invariant w.r.t. the addition of a constant value. 
% This might be one of the reasons why background level subtraction steps are not required for the convolutional neural networks.
We also consider \emph{occlusion maps}~\citep{Zeiler2014} for Net1(32,64). To determine if a certain area of an image is important for the classification, one can mask this area and see how this affects the prediction of the model. In Figure~\ref{fig:occlusion}, occlusion maps are shown for several \realsource and \bogussource instances. In these occlusion maps, the value of a pixel $(x,y)$ represents the predicted probability of the correct class by the model, after all pixels in a $3 \times 3$ square centred on pixel $(x,y)$ have been set to zero. Using the occlusion maps, one can gain insight into the way the model makes its predictions. For the \realsource sources, occluding pixels in the centre of an image causes misclassifications, while occluding the edges of the image has little to no effect. This is a good sign, as any other behaviour could indicate a reliance on artefacts or patterns in other regions of the image than the center. For the \bogussource instances, occluding any part of the image often makes little to no difference in classification. This indicates that the network  learns that the absence of a source is an indicator for \bogussource and as such, obscuring the \bogussource source will still result in a correct classification.

In summary, as expected, the shallow networks seem to mainly focus on the centre (\eg, ``is there anything in the centre in the \after image''). 
% Furthermore, they appear to be somewhat invariant against different background levels due to certain weight sums of the convolutional layers being very close to zero.

\subsubsection{Data Augmentation}

One of the main challenges that needs to be addressed is the shift from the given training data to completely new, unseen objects. Since the number of \realsource sources is very small (\eg, only very few distinct supernovae~objects are known and, hence, available for training), one has to develop a system that cannot only detect similar objects, but also new \realsource sources whose image representation is related, but different (\eg, a rotated version of an image containing a star, supernova, or artefact). A simple yet effective approach to improve the generalisation performance of convolutional neural networks is to augment the training data, see Section~\ref{sec:network_structures} for details and the particular augmentation steps conducted. Note that we only apply the augmentation on the training, the test set is not modified.

% \todot{Fabian: The following is already explained in section 3.3. Do we need this again or is there anything different written here? If not, please remove the overlap and refer to 3.3.}
%
%
% from here:
% A simple yet effective approach to improve the generalisation performance of convolutional neural networks is usually to \emph{augment} the training data. The general idea of such augmentation steps is to simulate possible candidates that are not found in the training data, but which can occur in new, unseen data. For our analysis, we apply the following two well-known operations on the training data:
% \begin{enumerate}
%  \item \emph{Rotations:} 
%  Depending on the particular setting, each of the training instances (all three images) are rotated by $\alpha \in \{90,95,100,180,185,190,270,275,280\}$ degrees. %NOTE: standard spline interpolation is done via scipy.ndimage.interpolation.rotate
%  \item \emph{Flippings:} Similarly, we consider flipping augmentation steps (vertical and horizontal mirroring).
% \end{enumerate}
% These augmentation steps are consecutively applied to the training data (\ie, rotations first followed by the flipping augmentations). 
% until here.
%
%
%  \todoAstro{No translation steps since real sources are centered. Details related to how rotations are done via scipy.} 
% \todot{Fabian: I think you explained this well in your answer.txt. Probably just write the same here?}
% We refer to the appendix for the data augmentation details and the involved parameters.

\begin{figure}
   \centering
   \hfill    
    \subfloat[Net2]{%
      \resizebox{0.4\columnwidth}{!}{\includegraphics{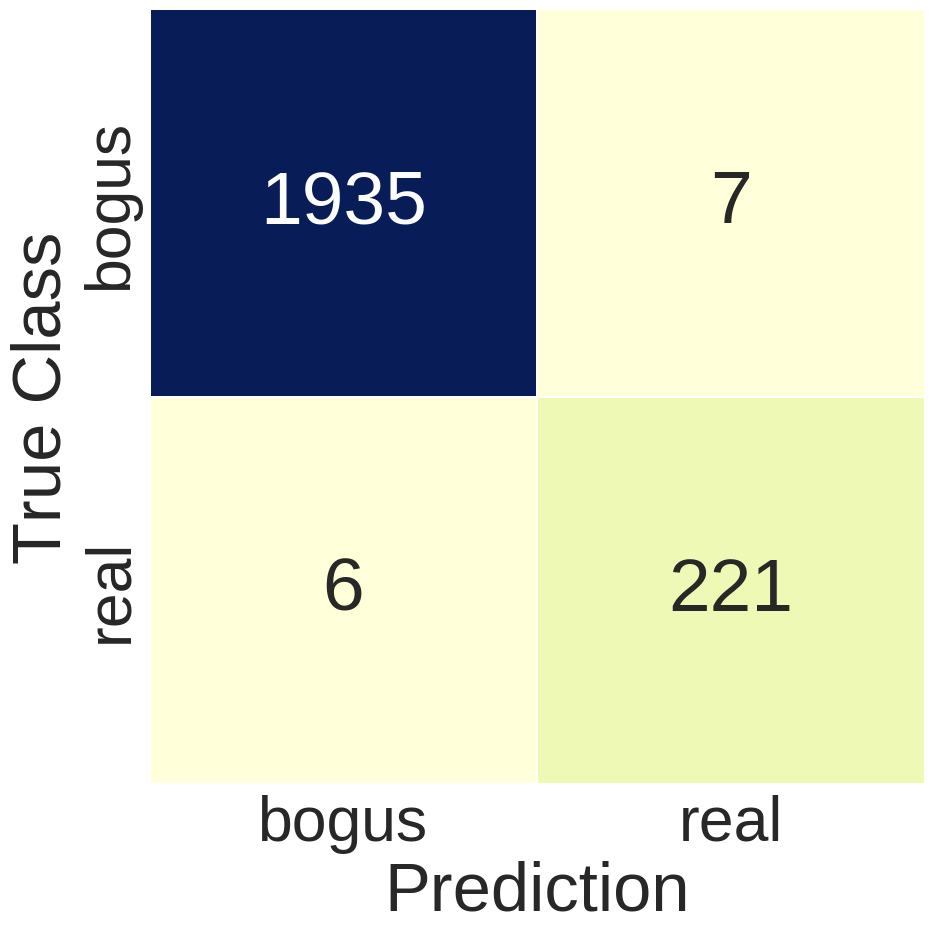}}
    }
    \hfill    
    \subfloat[Net3]{%
       \resizebox{0.4\columnwidth}{!}{\includegraphics{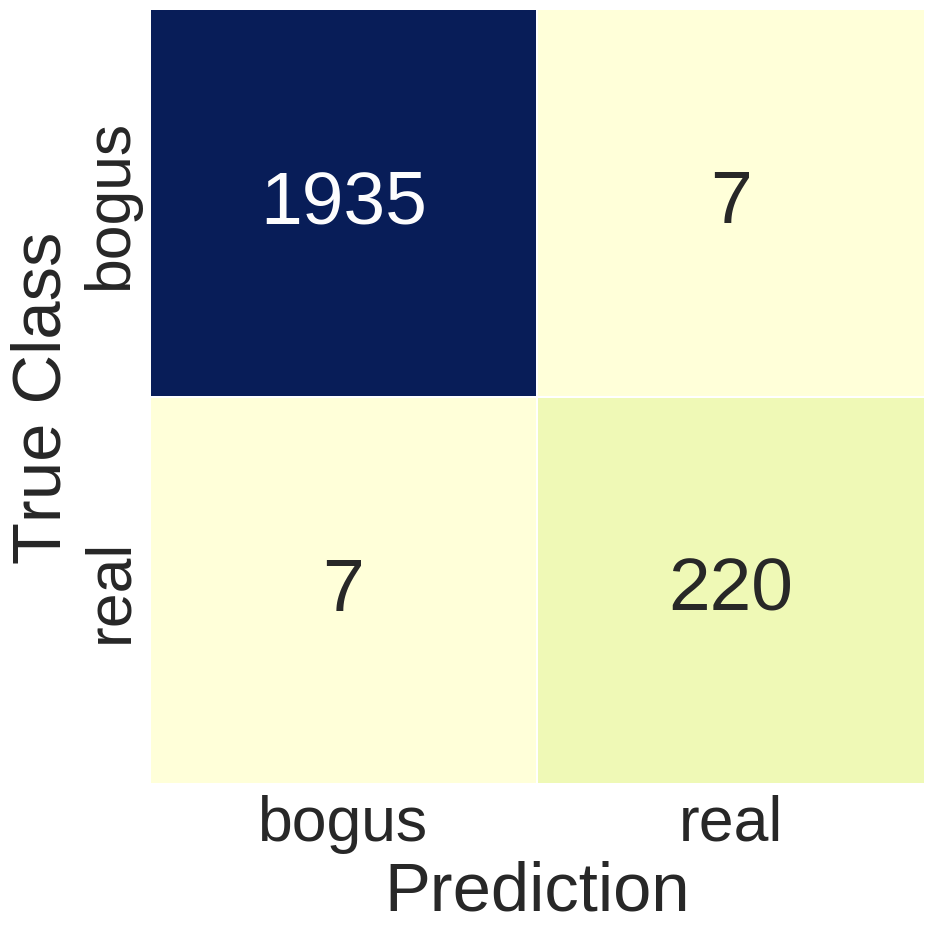}}
    }
    \hfill    
  \caption{Confusion matrices for Net2 and Net3 with data augmentation steps conducted in the preprocessing phase.}
  \label{fig:cnn_data_augmentation}
\end{figure}

The results for the deeper networks, Net2 and Net3, enhanced with the data augmentation steps are shown in Figure~\ref{fig:cnn_data_augmentation}. Given that we increase the number of instances in the training data with augmentation, we also increase the number of training iterations from 1,000 to 5,000. The confusion matrices in Figure~\ref{fig:cnn_data_augmentation} show that the addition of a data augmentation step can further improve the classification performance with Net3 only causing 14 overall misclassifications for the test dataset. We expect further data augmentations steps to be helpful as well in this context, see Section~\ref{sec:conclusions}.

\subsubsection{Analysis of Misclassifications}
\begin{figure}
  \centering
      \resizebox{0.45\textwidth}{!}{\includegraphics{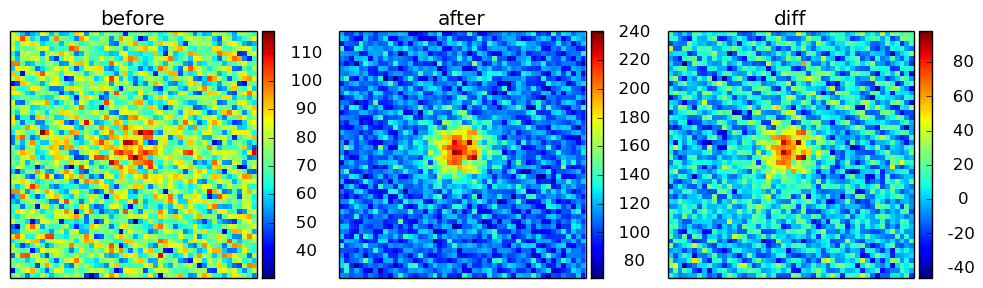}}
      \resizebox{0.45\textwidth}{!}{\includegraphics{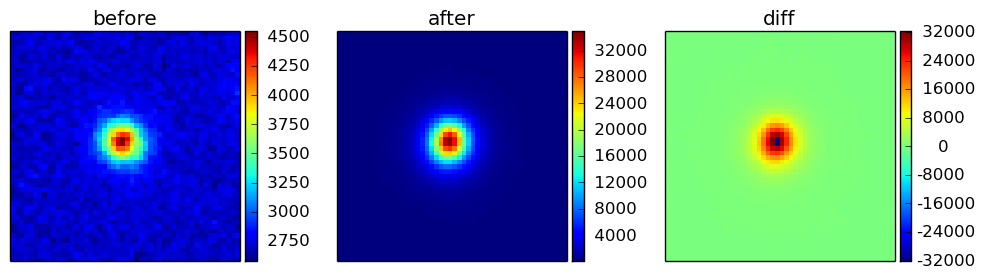}}
      \resizebox{0.45\textwidth}{!}{\includegraphics{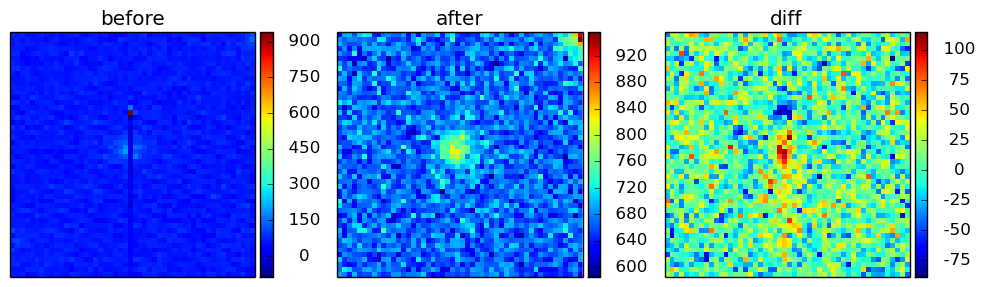}}
      \resizebox{0.45\textwidth}{!}{\includegraphics{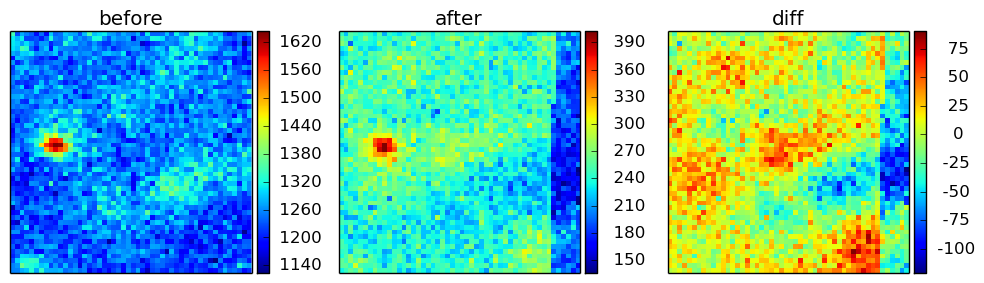}}
      \resizebox{0.45\textwidth}{!}{\includegraphics{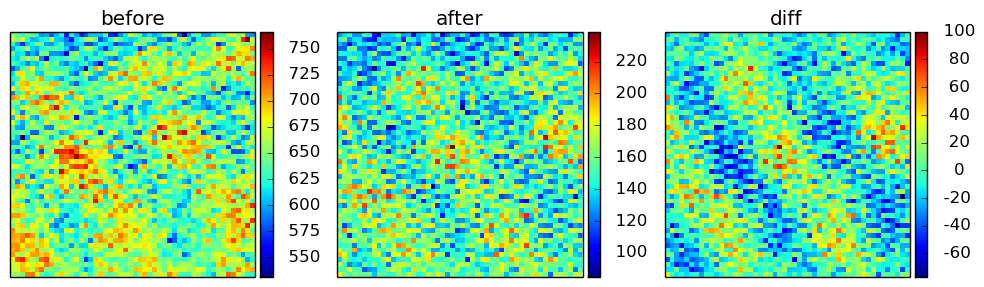}}
      \resizebox{0.45\textwidth}{!}{\includegraphics{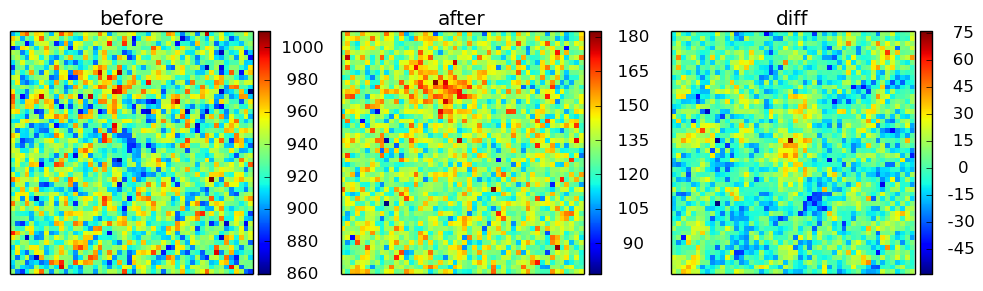}}
    \caption{Misclassifications made by Net3 with data augmentation (\bogussource instances misclassified as \realsource). The different colours along with the colour bars illustrate the pixel intensities per image.}
    \label{fig:misclassifications_net3_0}  
\end{figure}

\begin{figure}
  \centering
  \resizebox{0.45\textwidth}{!}{\includegraphics{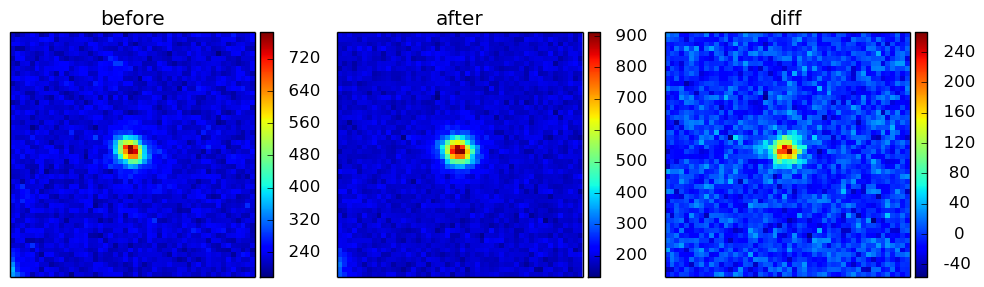}}
  \resizebox{0.45\textwidth}{!}{\includegraphics{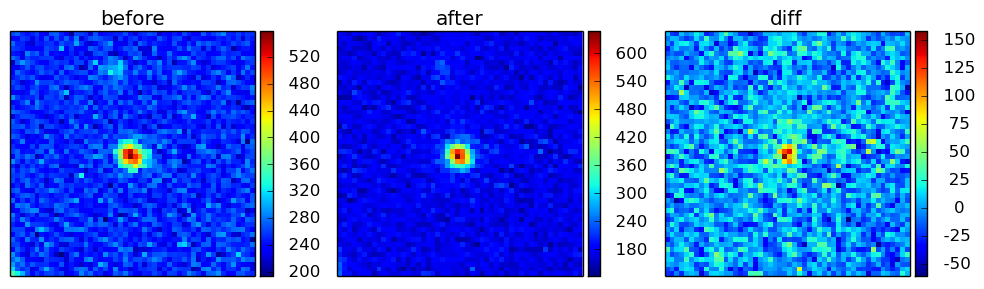}}
  \resizebox{0.45\textwidth}{!}{\includegraphics{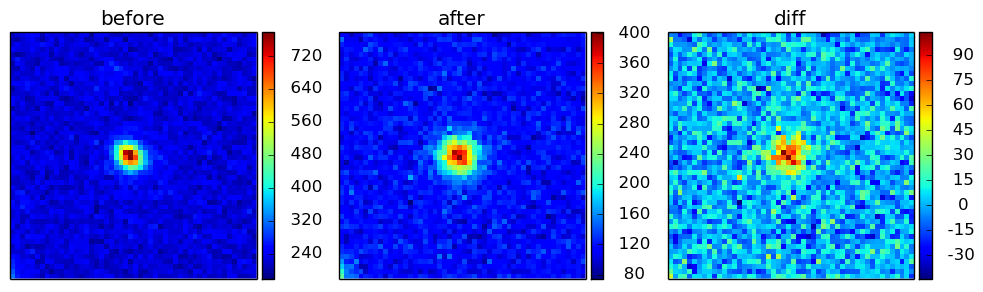}}
  \resizebox{0.45\textwidth}{!}{\includegraphics{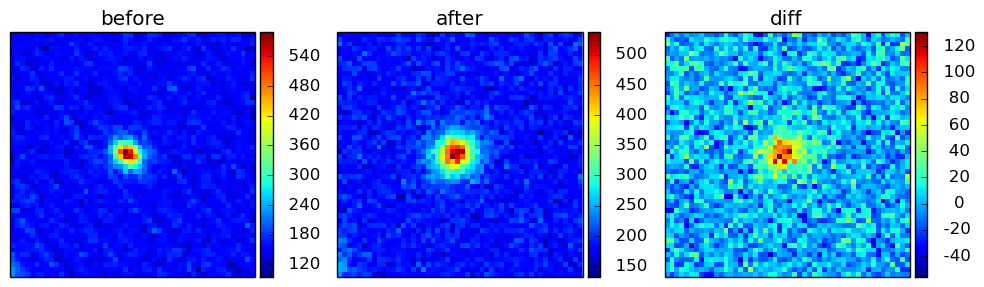}}
  \resizebox{0.45\textwidth}{!}{\includegraphics{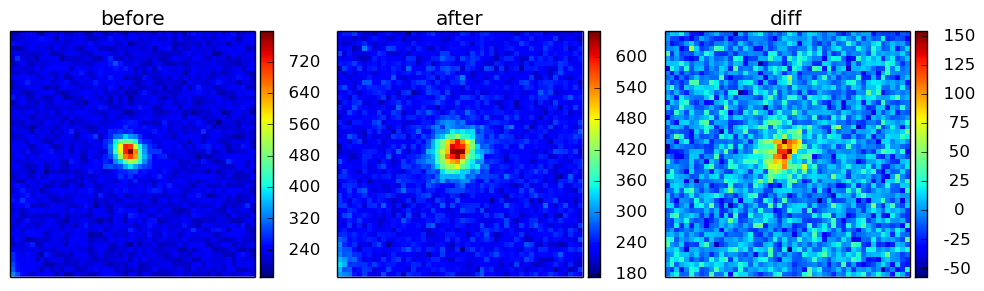}}
  \resizebox{0.45\textwidth}{!}{\includegraphics{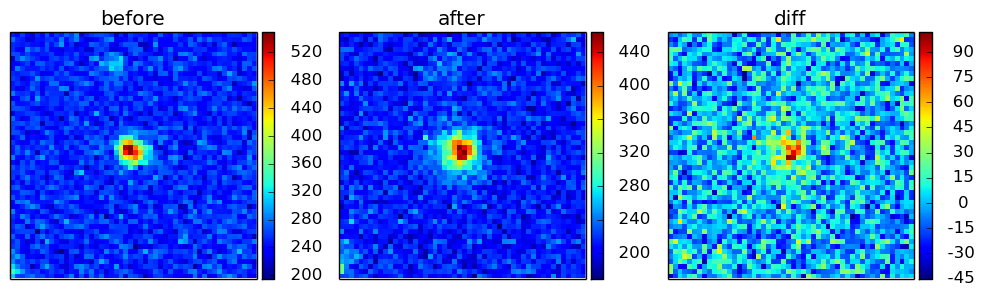}}
  \resizebox{0.45\textwidth}{!}{\includegraphics{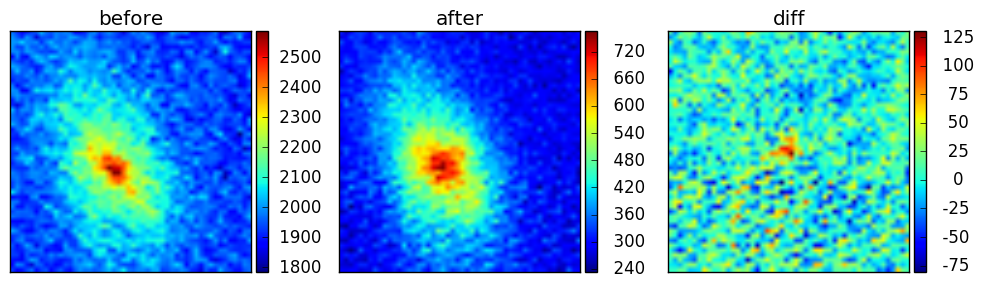}}
  \resizebox{0.45\textwidth}{!}{\includegraphics{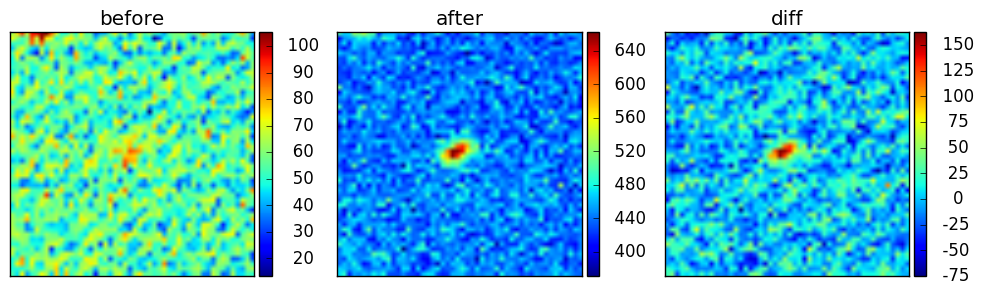}}
\caption{Misclassifications made by Net3 with data augmentation (\realsource instances misclassified as \bogussource). The different colours along with the colour bars illustrate the pixel intensities per image.}
\label{fig:misclassifications_net3_1}  
\end{figure}

Both Net2 and Net3 only misclassify a small number of test instances. In
Figures~\ref{fig:misclassifications_net3_0} and
\ref{fig:misclassifications_net3_1}, all misclassifications made by Net3 (with
data augmentation) are shown. For the first type of error, \bogussource
objects misclassified as \realsource, common examples are due to non-uniform
background noise in the template and/or target images, or deficits in the
template image that, after convolution, resemble point sources in the
difference image.  For the other type of error, \realsource instances
misclassified as \bogussource, a single object (with multiple observations) is
misclassified (top six images). This is very reasonable since it is generally
very hard for any model to distinguish varying stars and multiple follow-up
observations of a new supernova. However, from a practical perspective, such
cases can easily be handled by flagging such a source as \realsource the first
time it is observed (and correctly classified as \realsource). The remaining
two (last two rows) depict a low signal-to-noise detection of a faint
supernova near the core of its host galaxy, and an asteroid moving quickly
enough to show a trail in the target image.

The two deeper convolutional neural networks yield significantly less misclassifications as the baseline random forest approach (see the appendix for some misclassifications made by the random forest). Interestingly, the misclassifications differ slightly, \ie, the ones of Net3 do not form a subset of those misclassified by the random forest. We will see that this can actually be beneficial when combining the different classifiers.

\subsubsection{Less Input}

The networks considered so far are trained on all three input images that are available for each instance. By providing all the data, the networks can automatically determine which input images are important (see discussion above concerning the weights). A natural question is whether a competitive performance can also be achieved using less input data. We consider two settings: (1) Using only the \before and \after images and (2) using only the \diff image. Note that the latter setting usually forms the basis for other techniques that extract features from the \diff images only.

\begin{figure}
    \centering
    \subfloat[Net1(32,64)]{%
      \resizebox{0.4\columnwidth}{!}{\includegraphics{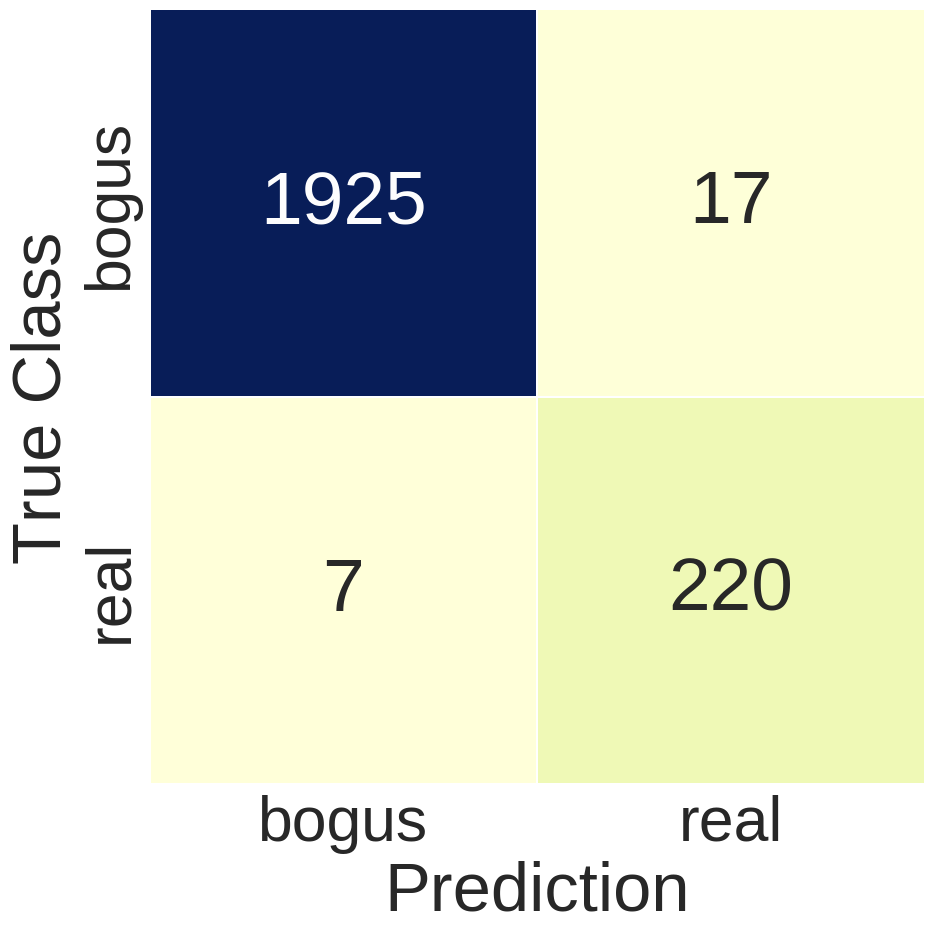}}
      \;\;\;\;\;\;\;\;
      \resizebox{0.4\columnwidth}{!}{\includegraphics{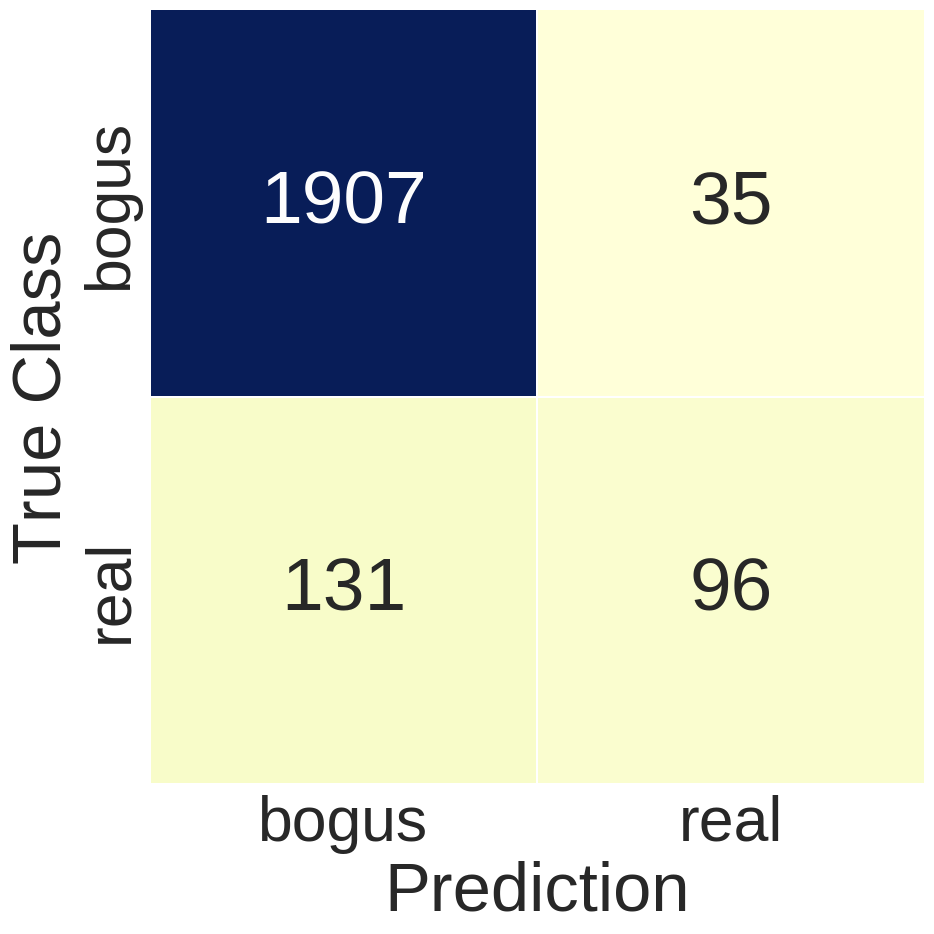}}
    }   	
    
    \subfloat[Net3]{%
      \resizebox{0.4\columnwidth}{!}{\includegraphics{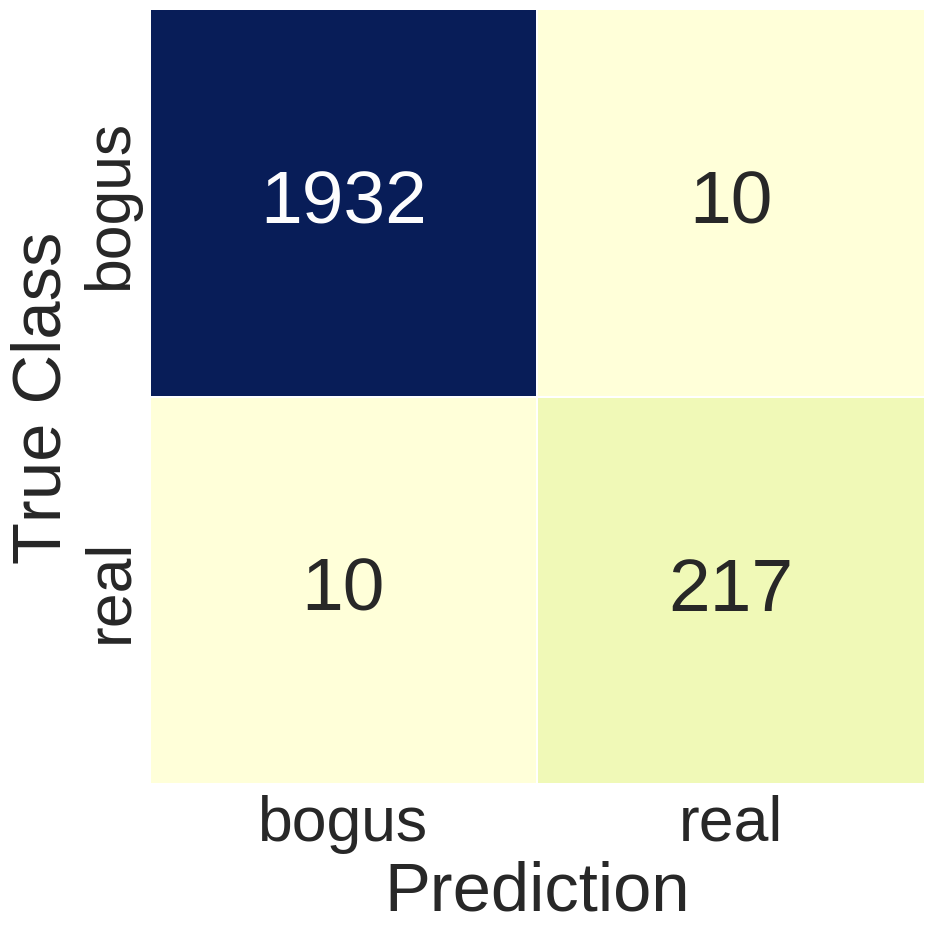}}
      \;\;\;\;\;\;\;\;
      \resizebox{0.4\columnwidth}{!}{\includegraphics{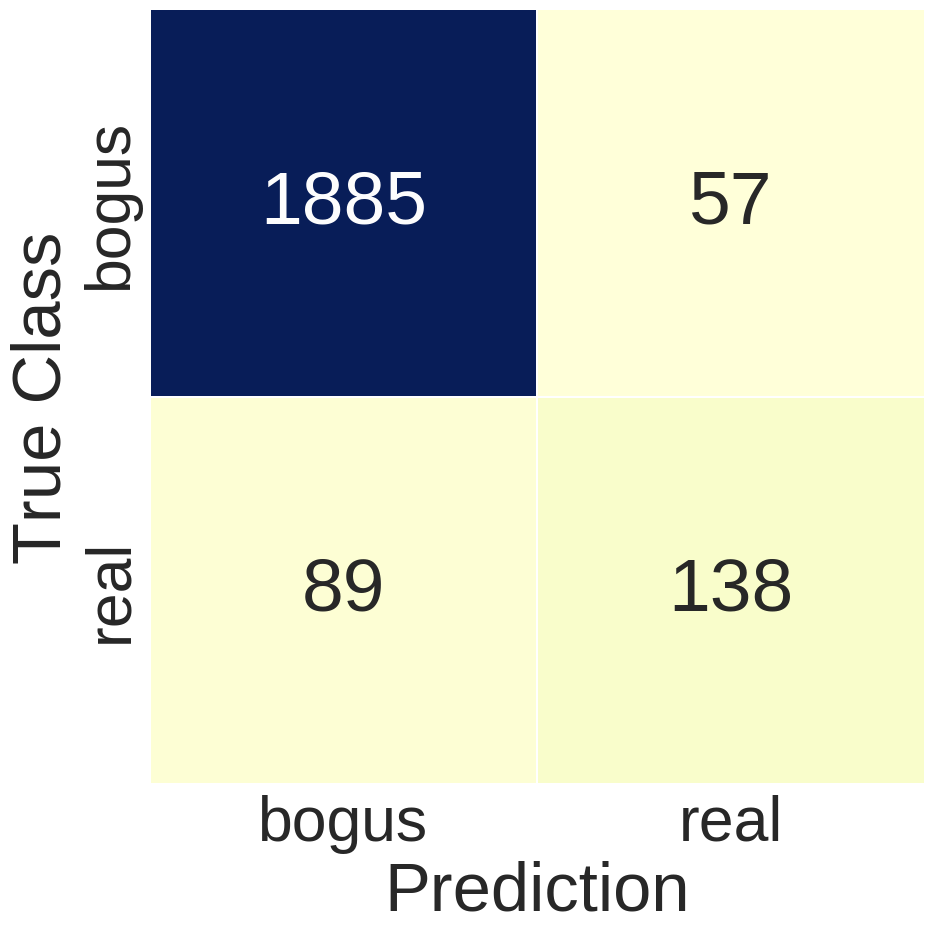}}
    }   	
  \caption{Classification performance of Net1(32,64) and Net3 (with data augmentation) in case only the \before and \after (left) or the \diff images  (right) are provided to the networks.}
  \label{fig:confusion_less_input}
\end{figure}

We focus on the simplest network considered in this work, Net1(32,64), and the best-performing one, Net3 with data augmentation. The induced confusion matrices are shown in Figure~\ref{fig:confusion_less_input}. By comparing this figure to Figure~\ref{fig:confusion}, it can be seen that using only \before and \after images yields a competitive performance compared to using all three input images. This may seem surprising due to the majority of the existing schemes being based on difference imaging. The results, however, clearly indicate that the reduced set of input images is sufficient for approaching the task at hand. This depicts a desirable outcome since one might be able to omit image subtraction steps in future detection pipelines. Further, the networks trained using only the \diff images yield a significantly worse classification performance. Hence, using only this type of information seems to be not enough for convolutional neural networks in this context.
% The performance is slightly better than before, which might be due to the less complex network structure (only two instead of three input images). 

\subsubsection{Ensembles}
\begin{figure}
    \centering
    \hfill
    \subfloat[E1]{%
      \resizebox{0.4\columnwidth}{!}{\includegraphics{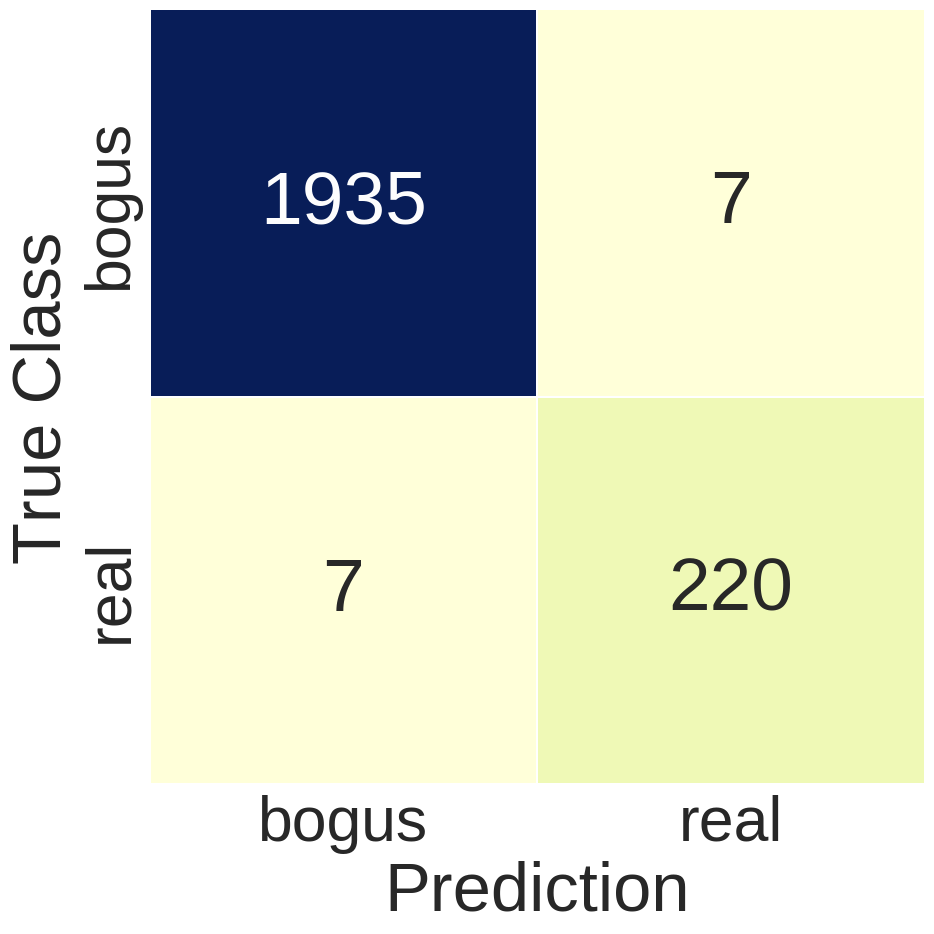}}
    }
    \hfill
    \subfloat[E2]{%
      \resizebox{0.4\columnwidth}{!}{\includegraphics{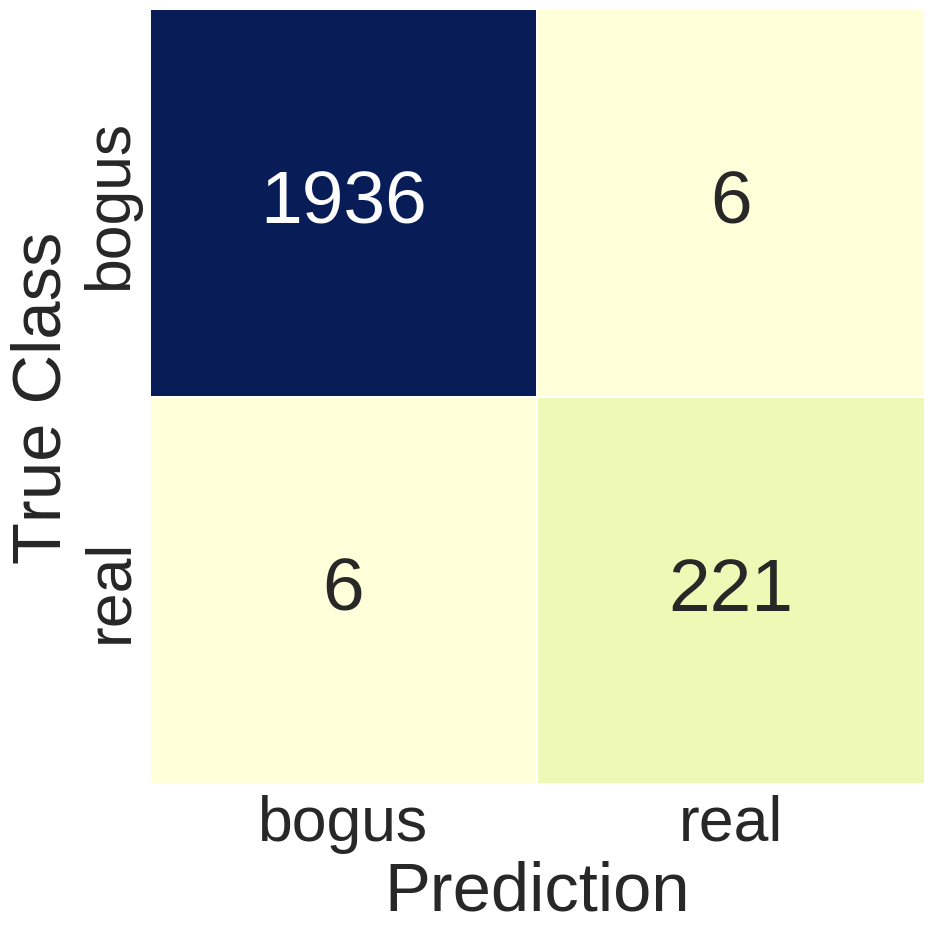}}
    }
    \hfill
  \caption{Confusion matrices for two ensemble classifiers. E1 combines three different neural networks. E2 two neural networks and a random forest.}
  \label{fig:ensembles}
\end{figure}

A common way to improve the classification performance is to consider ensembles of different models. As mentioned above, random forests depict ensembles of classification or regression trees and usually yield a significantly better performance than the individual models. We consider two ensembles E1 and E2: 
\begin{itemize}
 \item \emph{E1:} Net2 (data augmentation), Net3 (data augmentation), Net1(32,64) (\before and \after images only), and Net3 (data augmentation, \before and \after images only).
 \item \emph{E2:} Net2 (data augmentation), Net3 (data augmentation), and the random forest model.
\end{itemize}
The results are shown in Figure~\ref{fig:ensembles}. It can be seen that ensembling reduce the number of misclassifications. Furthermore, incorporating the random forest appears to be beneficial, potentially due to features that capture the characteristics of special cases. 

The improvements over the best-performing single convolutional neural networks are really small and, due to the relatively small test dataset, we do not argue that the ensembles outperform the individual classifiers. Nevertheless, the ensembles might exhibit a slightly better performance on completely new, unseen data since the combination of many different classifiers usually yield more ``stable'' results.

\subsubsection{Receiver Operating Characteristic (ROC) Analysis}

All results reported so far are based on the default threshold of $0.5$ for deciding which class an instance should belong to given the probability scores. For random forests, this simply corresponds to a majority vote among the individual trees. For the convolutional neural networks, it means the class with the highest probability. In general, many more \bogussource than \realsource instances are observed in practice and one might want to adapt the choice for the threshold. For example, one might prefer finding more \realsource sources at the cost of an increase in false positives (\ie, \bogussource instances misclassified as \realsource). This naturally depends on the number of human experts being available for manual inspection of all instances classified as \realsource~sources by the model.
% {\bf [I'm not sure we want this!] :)}

To quantify the performance of a model across a range of thresholds, one can make use of so-called \emph{receiver operating characteristic}~(ROC) curves~\citep{Fawcett:2006:IRA:1159473.1159475}. Here, the recall $tp \cdot (tp+fn)^{-1} = tp \cdot P^{-1}$ is also called \emph{true positive rate} (TPR), where $P$ denotes the number of all positive (\realsource) instances. Accordingly, one can define the \emph{false positive rate} (FPR) as $fp \cdot (fp+tn)^{-1}= fp \cdot N^{-1}$, where $N$ corresponds to all negative (\bogussource) objects. A classifier assigning only the class \realsource to all instances would therefore achieve an optimal TPR of 1.0, but also a potentially very large FPR. Ideally, one would like to have a large TPR and a small FPR; a ROC curve captures this trade-off.

\begin{figure}
\centering
 \resizebox{0.98\columnwidth}{!}{\includegraphics{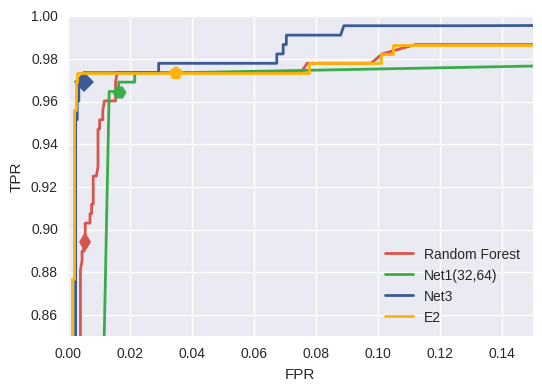}}
\caption{ROC curves for various models. The individual performances for a threshold of $0.5$ are marked for each curve.}
\label{fig:ROC}
\end{figure}

In Figure~\ref{fig:ROC} the ROC curves for various models are shown. Of the models plotted, Net3 has the best performance, which we can verify by calculating the \emph{area under the ROC curve} (AUC). The AUC values for the models are given in Table~\ref{AUC}. To test the significance of the differences in AUC values, we apply two statistical tests for ROC curves: the so-called \emph{DeLong}~\citep{Delong1988} and the \emph{bootstrap}~\citep{Hanley1983} methods. In both cases, we test the null hypothesis that the performance of both models is the same against the alternative hypothesis that Net3 performs better than the random forest (one-sided test). This results in \emph{p}-values of 0.0359 and 0.0352, respectively, indicating that Net3 has a significantly better performance than the state-of-the-art random forest approach. Even though this improvement seems small, it could result in a large decrease in false positives due to the large number of transient candidates that are generated each night.

The confusion matrix for NET1(32,64) shows that we have a TPR of $0.956$ for the standard threshold of 0.5. By moving right on the ROC curve, both the TPR and FPR increase and the decision of which FPR is still deemed acceptable is up to the user. In the case of transient vetting, the optimal threshold is determined by the capability to do follow-up studies on the possible transients and the willingness to search through a lot of extra \bogussource candidates to find a couple more transients. Such decisions have to be made per project and based on the human resources that are available to manually check the output of the processing pipelines.

\begin{table}
\centering
\caption{AUC for the different models.}
\label{AUC}
\begin{tabular}{ll} \hline
Model & AUC \\ \hline
Random forest  & 0.9907       \\
Net1(32,64)    & 0.9914       \\
Net3           & 0.9972       \\
E2             & 0.9946       \\ \hline
\end{tabular}
\end{table}

\section{Conclusions and Outlook}
\label{sec:conclusions}

We propose deep convolutional neural networks for the task of detecting astrophysical transients in future all-sky survey telescopes. The currently used state-of-the-art approach is based on feature extraction and a subsequent application of random forest algorithms. In our experimental evaluation, we demonstrate that even conceptually simple networks yield a competitive performance, which can be improved further via deeper architectures, data augmentation steps, and ensembling techniques. 
% We expect such systems to be useful for the upcoming large sky surveys. {\bf [this is already a repetition of the first sentence}] 
It is also worth mentioning that the networks considered also perform well (or even better) by just using template and target images, i.e., the networks do not rely on image subtraction. This might pave the way for future classification pipelines not containing image subtraction preprocessing steps.
% be important for the development of future detection pipelines.
% Further, the networks considered in this work can also successfully classify the objects at hand without relying on difference images, which might pave the way for future detection pipelines not containing image subtraction steps at all.

The machine learning models proposed in this work can be adapted and extended in various ways. Future telescope projects will produce significantly more data and we expect that taking such 
% {\bf [which ones, be more specific]} 
additional training instances into account will be beneficial to further improve the classification performances. The detection of extremely rare objects or artefacts will always depict a problem (even with better models due to many more objects being considered per night). Appropriate data preprocessing and augmentation steps conducted in the training phase might be one way to handle such instances correctly. In addition, adapting deep convolutional neural networks to the specific needs of the tasks at hand might be essential to cope with upcoming learning scenarios in this field (\eg, by considering specific loss functions that are suitable for extremely unbalanced datasets). We plan to investigate such important and interesting extensions in the near future.

\section*{Acknowledgements}

FG and VARMR acknowledge financial support from the Radboud Excellence Initiative. VARMR further acknowledges financial support from FCT in the form of an exploratory project of reference IF/00498/2015, from CIDMA strategic project UID/MAT/04106/2013, and from Enabling Green E-science for the Square Kilometer Array Research Infrastructure (ENGAGE SKA), POCI-01-0145-FEDER-022217, funded by Programa Operacional Competitividade e Internacionaliza\c{c}\~{a}o (COMPETE 2020) and FCT, Portugal.

% from ENgAGE SKA ROTEIRO/0041/2013/022217. Parts of this research were conducted by the Australian Research Council Centre of Excellence for All-sky Astrophysics (CAASTRO), through project number CE110001020.

%%%%%%%%%%%%%%%%%%%%%%%%%%%%%%%%%%%%%%%%%%%%%%%%%%



\bibliographystyle{mnras}

% \begin{thebibliography}{99}
% \bibitem[\protect\citeauthoryear{Author}{2012}]{Author2012}
% Author A.~N., 2013, Journal of Improbable Astronomy, 1, 1
% \bibitem[\protect\citeauthoryear{Others}{2013}]{Others2013}
% Others S., 2012, Journal of Interesting Stuff, 17, 198
% \end{thebibliography}

\bibliography{biblio}

%%%%%%%%%%%%%%%%%%%%%%%%%%%%%%%%%%%%%%%%%%%%%%%%%%

%%%%%%%%%%%%%%%%% APPENDICES %%%%%%%%%%%%%%%%%%%%%

\appendix

% \section{Implementation}
% \label{appendix:implementation}

\section{Misclassifications}
\begin{figure}
  \centering
      \resizebox{0.44\textwidth}{!}{\includegraphics{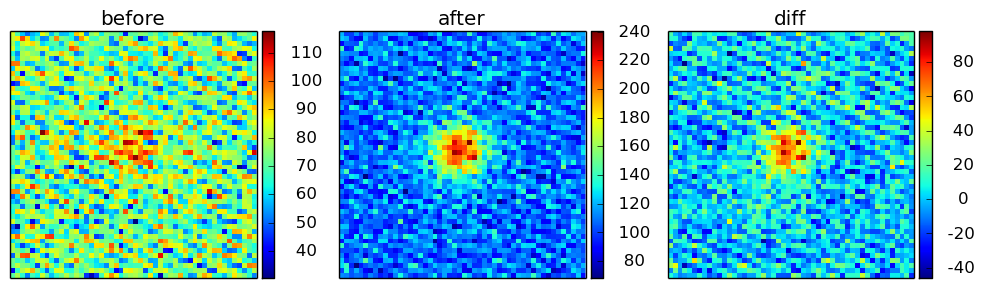}}
      \resizebox{0.44\textwidth}{!}{\includegraphics{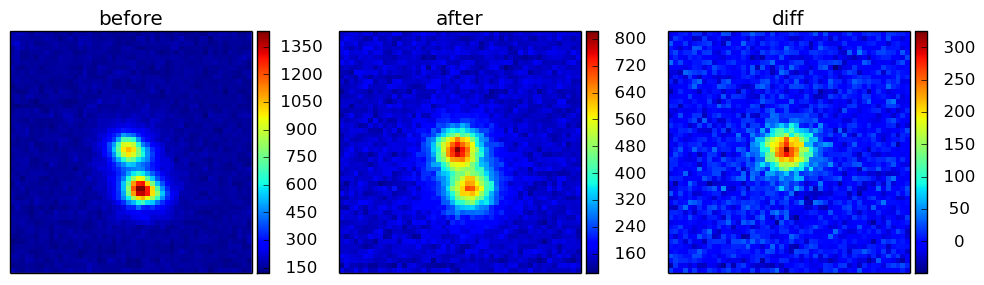}}
      \resizebox{0.44\textwidth}{!}{\includegraphics{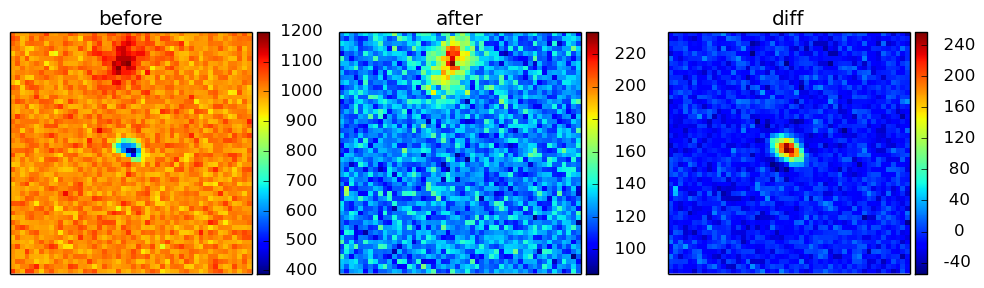}}
      \resizebox{0.44\textwidth}{!}{\includegraphics{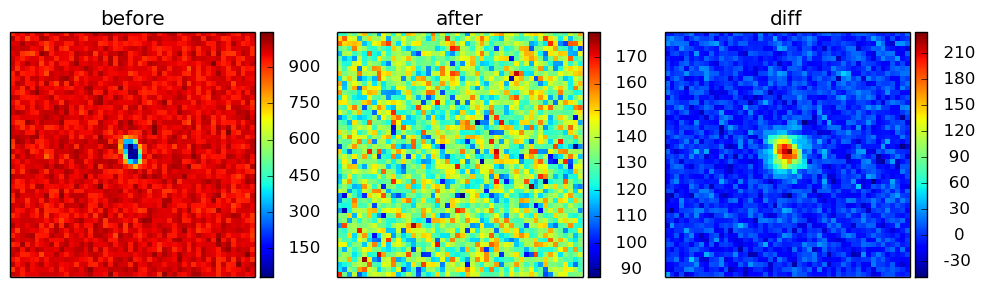}}
      \resizebox{0.44\textwidth}{!}{\includegraphics{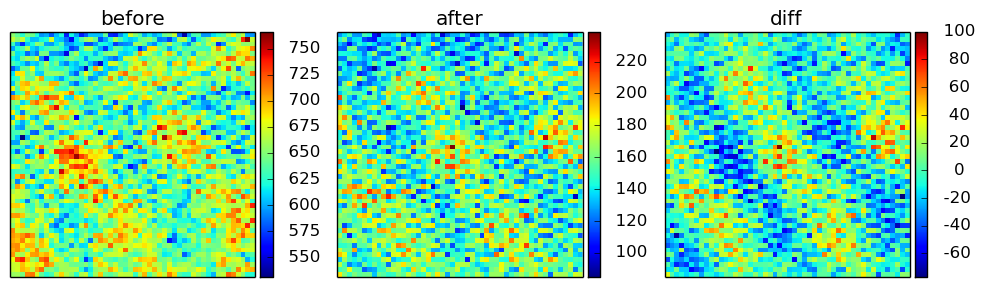}}
      \resizebox{0.44\textwidth}{!}{\includegraphics{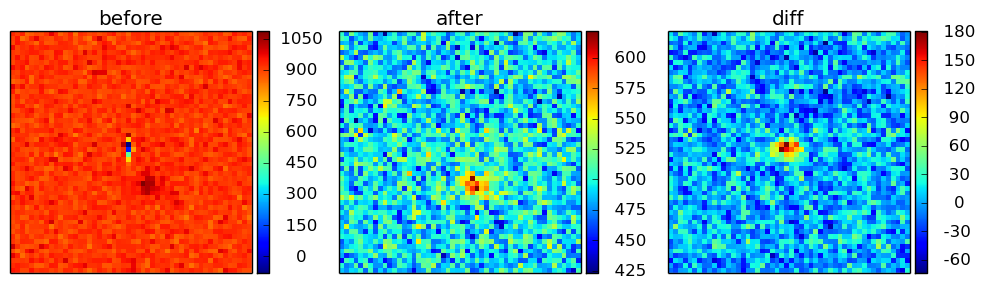}}
      \resizebox{0.44\textwidth}{!}{\includegraphics{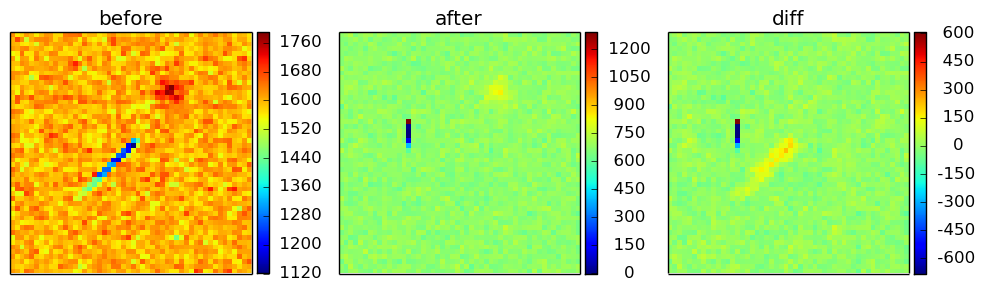}}
      \resizebox{0.44\textwidth}{!}{\includegraphics{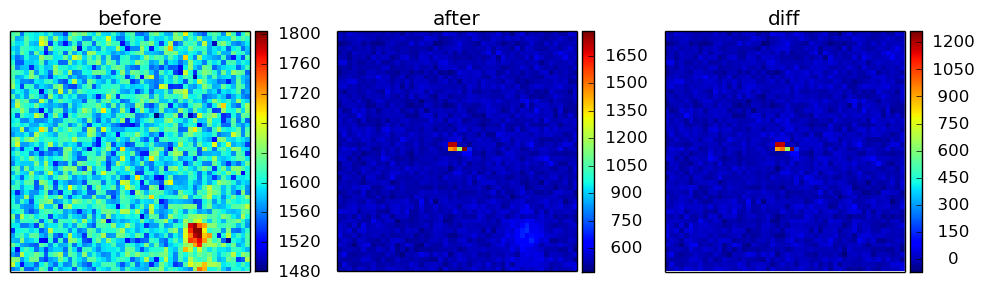}}
      \resizebox{0.44\textwidth}{!}{\includegraphics{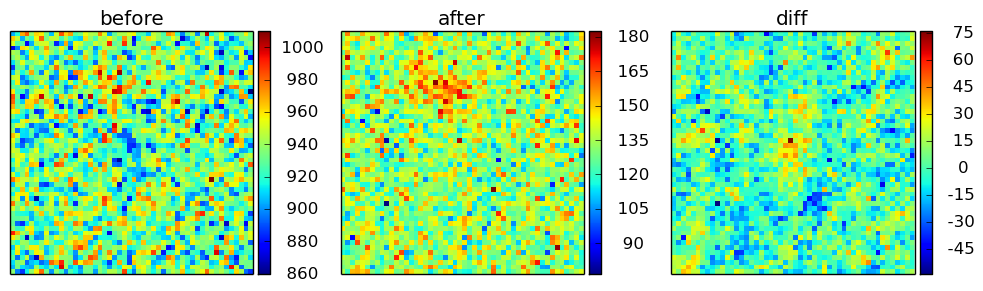}}
      \resizebox{0.44\textwidth}{!}{\includegraphics{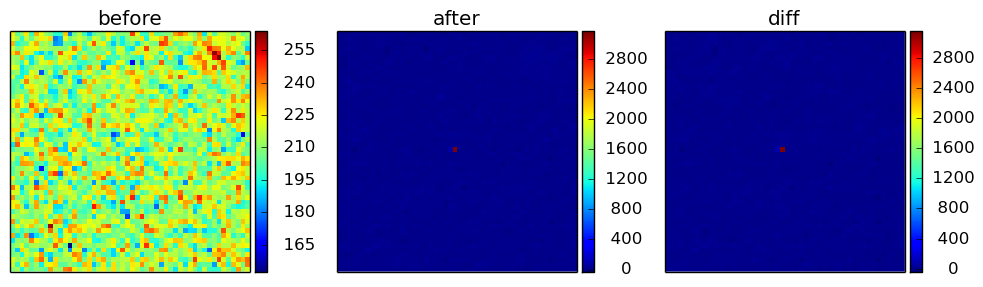}}
    \caption{A subset of the \bogussource objects misclassified as \realsource by the random forest model. The different colours along with the colour bars illustrate the pixel intensities per image.}
    \label{fig:misclassifications_forest_0}  
\end{figure}

\begin{figure}
  \centering
  \resizebox{0.44\textwidth}{!}{\includegraphics{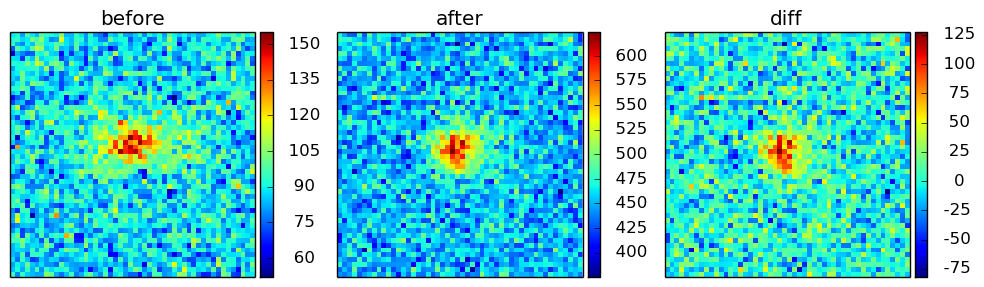}}
%   \resizebox{0.45\textwidth}{!}{\includegraphics{images/experiments/misclassifications/forest/1/4.png}}
%   \resizebox{0.45\textwidth}{!}{\includegraphics{images/experiments/misclassifications/forest/1/5.png}}
  \resizebox{0.44\textwidth}{!}{\includegraphics{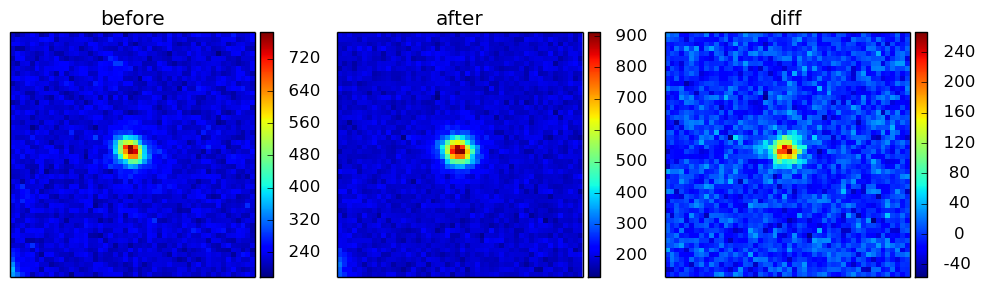}}
%   \resizebox{0.45\textwidth}{!}{\includegraphics{images/experiments/misclassifications/forest/1/29.png}}
%   \resizebox{0.45\textwidth}{!}{\includegraphics{images/experiments/misclassifications/forest/1/30.png}}
%   \resizebox{0.45\textwidth}{!}{\includegraphics{images/experiments/misclassifications/forest/1/31.png}}
%   \resizebox{0.45\textwidth}{!}{\includegraphics{images/experiments/misclassifications/forest/1/32.png}}
%   \resizebox{0.45\textwidth}{!}{\includegraphics{images/experiments/misclassifications/forest/1/33.png}}
  \resizebox{0.44\textwidth}{!}{\includegraphics{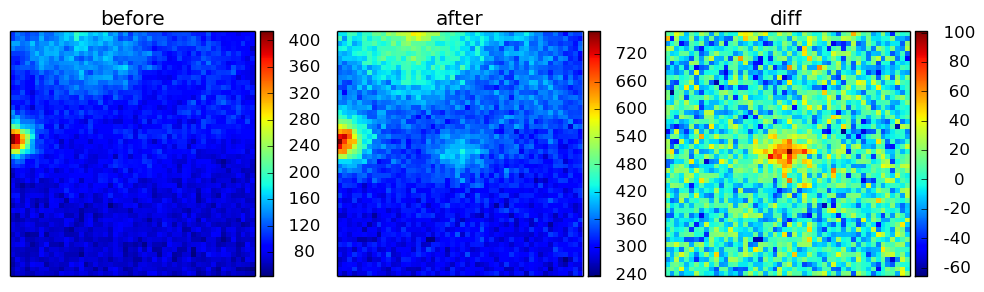}}
%   \resizebox{0.45\textwidth}{!}{\includegraphics{images/experiments/misclassifications/forest/1/52.png}}
%   \resizebox{0.45\textwidth}{!}{\includegraphics{images/experiments/misclassifications/forest/1/53.png}}
%   \resizebox{0.45\textwidth}{!}{\includegraphics{images/experiments/misclassifications/forest/1/58.png}}
%   \resizebox{0.45\textwidth}{!}{\includegraphics{images/experiments/misclassifications/forest/1/59.png}}
  \resizebox{0.44\textwidth}{!}{\includegraphics{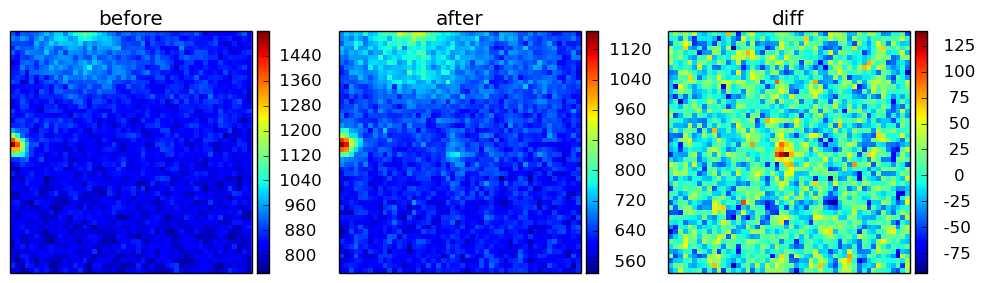}}
%   \resizebox{0.45\textwidth}{!}{\includegraphics{images/experiments/misclassifications/forest/1/61.png}}
%   \resizebox{0.45\textwidth}{!}{\includegraphics{images/experiments/misclassifications/forest/1/62.png}}
%   \resizebox{0.45\textwidth}{!}{\includegraphics{images/experiments/misclassifications/forest/1/65.png}}
  \resizebox{0.44\textwidth}{!}{\includegraphics{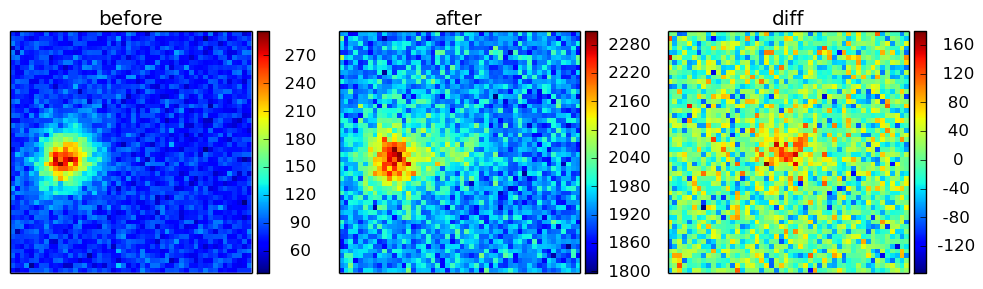}}
  \resizebox{0.44\textwidth}{!}{\includegraphics{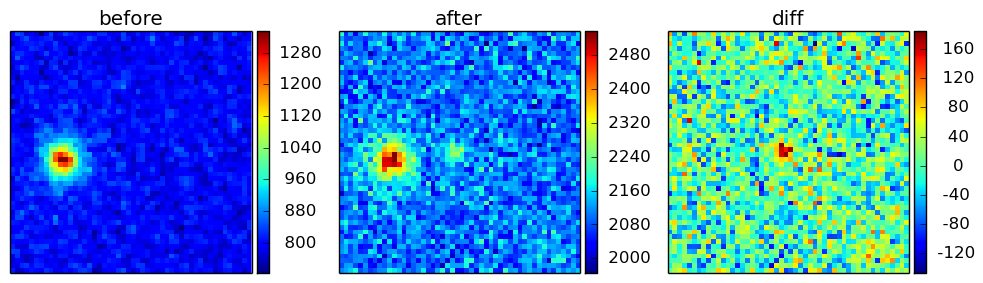}}
  \resizebox{0.44\textwidth}{!}{\includegraphics{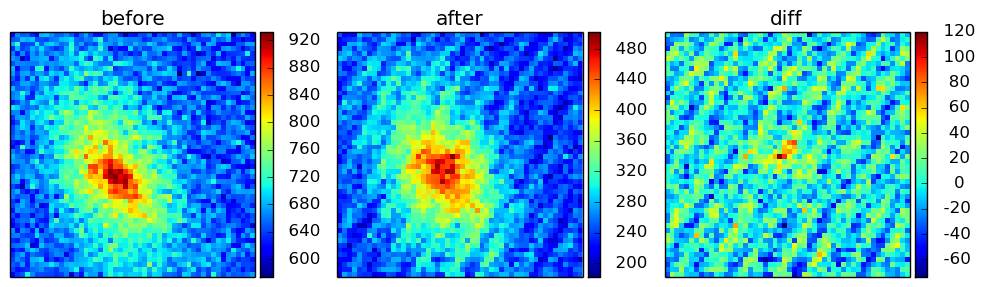}}
  \resizebox{0.44\textwidth}{!}{\includegraphics{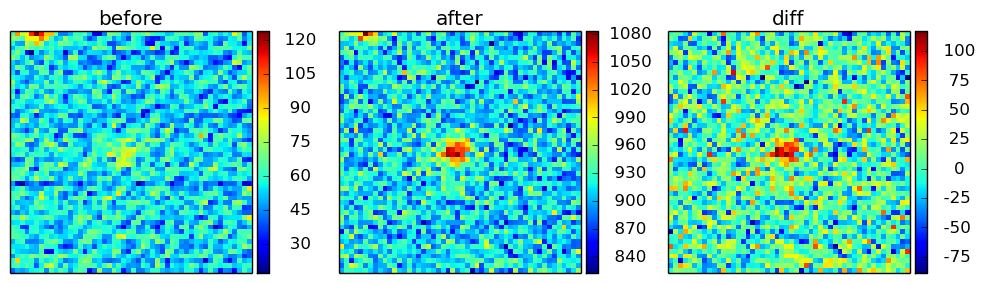}}
  \resizebox{0.44\textwidth}{!}{\includegraphics{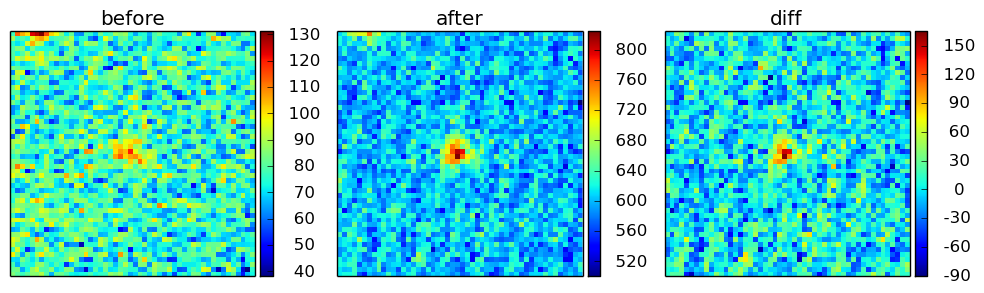}}
  \resizebox{0.44\textwidth}{!}{\includegraphics{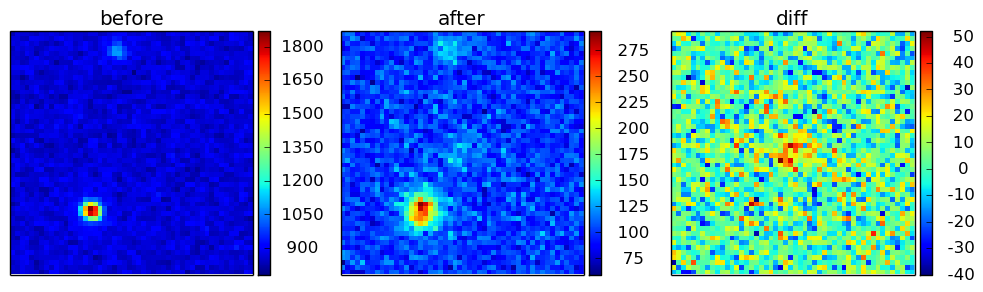}}
\caption{A subset of the \realsource objects misclassified as \bogussource by the random forest model. The different colours along with the colour bars illustrate the pixel intensities per image.}
\label{fig:misclassifications_forest_1}  
\end{figure}

Figures~\ref{fig:misclassifications_forest_0} and \ref{fig:misclassifications_forest_1} show misclassifications made by the random forest baseline. All false positives instances are given in Figure~\ref{fig:misclassifications_forest_0}, whereas only a subset is given for the false negatives in Figure~\ref{fig:misclassifications_forest_1}.

%%%%%%%%%%%%%%%%%%%%%%%%%%%%%%%%%%%%%%%%%%%%%%%%%%

% Don't change these lines
\bsp	% typesetting comment
\label{lastpage}
\end{document}